\documentclass[aps,prd,twocolumn,amsmath,amssymb,showpacs,floatfix,superscriptaddress,nofootinbib]{revtex4-1}

\pdfoutput=1
\usepackage{natbib}
\usepackage[utf8]{inputenc} 
\usepackage{amsmath}
\usepackage{amssymb}
\usepackage{bm}
\usepackage{latexsym}
\usepackage{graphicx,epstopdf}
\usepackage{epstopdf}
\DeclareGraphicsExtensions{.ps, .png}
\usepackage{dcolumn} 
\usepackage{footnote}
\usepackage{tabularx,ragged2e,booktabs}
\usepackage{breqn}
\usepackage[normalem]{ulem}
\usepackage{float}
\restylefloat{table}
\usepackage{color}

\newcommand{\refsec}[1]{section~\ref{sec:#1}}
\newcommand{\refeq}[1]{Eq.~(\ref{eq:#1})}

\newcommand{\reffig}[1]{Fig.~\ref{fig:#1}}

\newcommand{\xef}{x_e^{\rm fid}}

\newcommand{\zmax}{z_{\rm max}}
\newcommand{\zmin}{z_{\rm min}}
\newcommand{\zre}{z_{\rm re}}
\newcommand{\xemin}{x_e^{\rm min}}

\newcommand{\tauhi}{\tau_{\rm hi}}
\newcommand{\taulo}{\tau_{\rm lo}}
\newcommand{\sample}{{\rm sample}}

\def\lsim{\mathrel{\raise.3ex\hbox{$$<$$\kern-.75em\lower1ex\hbox{$\sim$}}}}
\def\gsim{\mathrel{\raise.3ex\hbox{$$>$$\kern-.75em\lower1ex\hbox{$\sim$}}}}

\newcommand{\beq}{\begin{equation}}
\newcommand{\eeq}{\end{equation}}

\newcommand{\bea}{\begin{eqnarray}}
\newcommand{\eea}{\end{eqnarray}}

\definecolor{darkgreen}{rgb}{0.0, 0.2, 0.13}
\newcommand{\todo}[1]{\textcolor{darkgreen}{\bf{#1}}}

\newcommand{\relike}{RELIKE}
\newcommand{\cosmomcrelike}{\texttt{CosmoMC-RELIKE}}

\definecolor{darkgreen}{cmyk}{0.85,0.2,1.00,0.2} 
\definecolor{purple}{cmyk}{0.5,1.0,0,0}

\definecolor{ultramarine}{rgb}{0.07, 0.04, 0.56}
\definecolor{cadmiumgreen}{rgb}{0.0, 0.42, 0.24}
\definecolor{indigo(dye)}{rgb}{0.0, 0.25, 0.42}
\usepackage[linktocpage=true]{hyperref}
\hypersetup{
colorlinks=true,
citecolor=ultramarine,
linkcolor=cadmiumgreen,
urlcolor=indigo(dye),
pdfauthor={},
pdftitle={},
pdfsubject={}
}

\begin{document}
	
\title{RELIKE: Reionization Effective Likelihood from Planck 2018 Data} 

\author{Chen Heinrich}\email{chenhe@caltech.edu}
\affiliation{$California\ Institute\ of\ Technology,\ Pasadena,\ California\ 91109,\ USA$}

\author{Wayne Hu}
\affiliation{$Kavli\ Institute\ for\ Cosmological\ Physics,\ Enrico Fermi Institute,\ University of Chicago,\ Chicago\ Illinois\ 60637$}
\affiliation{$Department\ of\ Astronomy\ \&\ Astrophysics,\ University\ of\ Chicago,\ Illinois\ 60637$}

\begin{abstract}

We release RELIKE (Reionization Effective Likelihood), a fast and accurate effective likelihood code based on the latest Planck 2018 data 
that allows one constrain any model for reionization between $6 < z < 30$ using five constraints from the CMB reionization principal components (PC). We tested the code on two example models which showed excellent agreement with sampling the exact Planck likelihoods using either a simple Gaussian PC likelihood or its full kernel density estimate.
This code enables a fast and consistent means for 
combining Planck constraints with other reionization data sets, such as kinetic Sunyaev-Zeldovich effects, line-intensity mapping, luminosity function, star formation history, quasar spectra, etc, where the redshift dependence of the ionization history is important.  
Since the PC technique tests any reionization history in the given range, we also derive model-independent constraints for the total Thomson optical depth $\tau_{\rm PC} = 0.0619^{+0.0056}_{-0.0068}$ and its $15\le z \le 30$ high redshift component $\tau_{\rm PC}(15, 30) < 0.020 $ (95\% C.L.).
The upper limits on the high-redshift optical depth is a factor of $\sim3$ larger than those reported in the Planck 2018 cosmological parameter paper using the FlexKnot method
and we validate our results with a direct analysis of a two-step model which permits this small high-$z$ component.

\end{abstract}
%\pacs{}

\maketitle

\section{Introduction}
\label{sec:intro}

The cosmic microwave background (CMB) has helped establish the $\Lambda$CDM model as the standard cosmological model~\cite{Aghanim:2018eyx}. While many components of this standard cosmology model are well-understood, the details of the process of reionization however, remains one of the most uncertain pieces (see e.g.~\cite{2016ASSL..423.....M} and references therein).  It is usually characterized by a single parameter, the total Thomson optical depth, whose  uncertainty propagates into the inferences of other important parameters such as the primordial power spectrum amplitude. Through it, the uncertainty in reionization will become one of the major sources of uncertainty for measuring the sum of neutrino masses from future gravitational lensing measurements of the CMB~ \cite{Smith:2006nk}, 
%Allison:2015qca
and will also have implications for inferring cosmic acceleration through the growth of structure~\cite{Hu:2003pt}. 

Many probes of  the reionization era exist today and with future probes it will only become more important to combine them consistently the full implications of the CMB. In the CMB data, free electrons during the reionization epoch Compton scatter with the CMB photons to cause a suppression of the CMB primary anisotropies; on the large scales, the same scattering process of the CMB temperature quadrupole in the frame of the free electrons induces an additional reionization bump in the CMB $E$-mode polarization. On the small scales, the kinetic Sunyaev-Zeldovich (kSZ) effects generated by the scattering of the CMB photons off ionized gas in bulk motion~\cite{Sunyaev:1970er, Sunyaev:1980vz} are used to constrain morphology and duration of reionization~\cite{mcquinn_2005,mesinger_2012_kSZ}. Additionally, observations of galaxy luminosity functions, quasars, star formation rate, line-intensity mapping also provide important and complementary information on reionization~\cite{2016ASSL..423.....M}. In this paper, we aim to extract all the information in the large-scale $E$-mode observations for constraining the global ionization history of the Universe, and provide a way to consistently analyze them with all of these other probes.
 
In the standard CMB analysis, reionization has been  modeled as a steplike transition in the global ionization history in the form of a tanh function, with the step location parameterized by the total Thomson optical depth induced. This steplike tanh model assumes, by construction, that there is negligible ionization before the transition. However, the shape of the reionization bump induced in the CMB $E$-mode polarization at large angles contains more information on the coarse-grained evolution of the ionization history than the total optical depth~\cite{Hu:2003gh, Mortonson:2007hq}. 

To extract the most information possible from this $E$-mode bump, Ref.~\cite{Hu:2003gh, Mortonson:2007hq} developed the principal component (PC) method, where a few PCs is sufficient to describe the entire model space of physical ionization histories regarding their observable impact on the large-angle $E$-mode power spectrum. This method has been applied to WMAP and Planck data to obtain complete constraints on reionization models~\cite{Mortonson:2008rx, Mortonson:2007hq, Heinrich:2016ojb, Aghanim:2018eyx}. It was also adopted for a Planck 2013 analysis for marginalizing ionization history when constraining inflationary parameters in Ref.~\cite{Planck:2013jfk}, for studying inflationary features vs reionization features in Ref.~\cite{Obied:2018qdr}, as well as massive neutrinos and gravitational waves in Ref.~\cite{Dai:2015dwa}. %\cite{Obied:2017tpd} 
In a re-analysis of the Planck 2015 data with PCs~\cite{Heinrich:2016ojb}, a component of the high-redshift ionization that was missed by a simple steplike model was mildly preferred.

In the latest official Planck 2018 release~\cite{Aghanim:2018eyx},  this mild preference for finite high redshift ionization in the Planck 2015 release turned into upper limits, largely due to the reduction of systematics at large-scales~\cite{Aghanim:2018eyx, Millea:2018bko} resulting in reduced $C_l^{EE}$ around $l \lesssim 10$ (see also Ref.~\cite{Heinrich:2018btc}). The FlexKnot method was primarily employed, which is also able to capture general ionization histories by varying the number of ``knots" in redshifts and the amplitude of the ionization fraction at these knots. 

Since the release of the Planck 2018 official analysis, an improved likelihood for the low-$\ell$ $E$-mode polarization was publicly released in 2019. This new likelihood,
$\texttt{SRoll2}$\footnote{SRoll2: \url{http://sroll20.ias.u-psud.fr.}}, adopts better foreground modeling techniques and allowing for improved reionization constraints \cite{Delouis:2019bub}.

Because the Planck data sets the current standard for CMB constraints, it is important to extract all the information present on reionization  so that it may be readily used in future joint analyses with other cosmological probes. In this paper, we obtain new reionization PC constraints using legacy Planck 2018 likelihood and this latest likelihood $\texttt{SRoll2}$ for the low-$\ell$ $E$-mode power spectrum. 

Enabled by the completeness property of the PCs, we turn these constraints into an effective means of  assessing the CMB likelihood of \textit{any} reionization model out to $\zmax$ = 30. 
The 95\% C.L. upper limit on the cumulative optical depth between $z=15$ and 30 is $\tau(15, 30) < 0.020$  and about $\sim3$ times less stringent than reported using FlexKnot in the official Planck analysis $\tau(15, 30)<0.007$~\cite{Aghanim:2018eyx}.  We validate our results with an explicit toy model that allows such a small component of high redshift ionization and show that this result is not simply a consequence of principal component priors alone as suggested by \cite{Millea:2018bko} nor should these priors be changed as was adopted in the official Planck analysis~\cite{Aghanim:2018eyx}.
Extending the range of PCs to cover up to $z_{\rm max} = 50$, we verify that our results are robust to this extension.

With the technique  validated,
we release the Reionization Effective LIKElihood code (\relike) which is available on GitHub at \url{https://github.com/chenheinrich/RELIKE}. There are two packages that come with this code release: 1)  a python package \relike, to quickly evaluate the effective Planck likelihood of any reionization model given $x_e(z)$. If desired, this package can be connected to a MCMC sampler to sample the posteriors in a higher-dimensional parameter space. 2)  the CosmoMC sampler with a fortran implementation of \relike\  code in the package \cosmomcrelike. This code enables a  much simpler and faster testing of specific reionization models with the Planck data, and provides a  consistent way of combining this information with other reionization datasets. 

The paper is structured as follows. We first review in \S\ref{sec:background} the methodology of  reionization principal components along with the kernel density estimate (KDE) technique and the Gaussian approximation \cite{Heinrich:2016ojb} used for building the effective likelihoods. Then in \S\ref{sec:results}, we describe the PC results using the Planck 2018 + \texttt{SRoll2} likelihoods, and compare with previous results in literature. In \S\ref{sec:effective_likelihood}, we test the effective likelihood code and demonstrate its fidelity with two examples: 1) the standard steplike model; and 2) a two-parameter model where a plateau of ionization at high-$z$ is added to the standard tanh model for illustration purposes. Finally, we summarize our results and conclude in \S\ref{sec:conclusion}.

\section{PC Technique}
\label{sec:background}
The principal component technique for constraining reionization using the large-angle $C_\ell^{EE}$ polarization spectrum was first introduced in Ref.~\cite{Hu:2003gh}. Here, we  briefly summarize the PC technique itself in \S\ref{sec:PC}. We present  techniques for building its effective likelihoods which can be used to rapidly explore any reionization model in \S\ref{sec:KDE}. We refer the readers to Refs.~\cite{Mortonson:2007hq,Heinrich:2016ojb} respectively for a more complete description of these topics.

\subsection{Reionization Principal Components}
\label{sec:PC}
We begin by parametrizing $x_e(z)$, the ionization fraction relative to the fully ionized hydrogen at redshift $z$, into its principal components $S_{a}(z)$ with respect to the CMB $E$-mode polarization:
\begin{equation}
x_e(z)=\xef(z)+\sum_{a}m_{a}S_{a}(z),
\label{eq:mmutoxe}
\end{equation}
where $m_a$ are the PC amplitudes and $\xef(z)$ is the fiducial model. We obtain the PC basis functions $S_{a}(z)$ as eigenfunctions of the Fisher information matrix for $x_e(z)$ in a given range $z_{\rm min}<z<z_{\rm max}$ from cosmic variance limited $C_\ell^{EE}$ measurements 
\beq
F_{ij} = \sum_l \frac{2 l+1}{2} \frac{\partial \ln C_l^{EE}}{\partial x_e(z_i)}\frac{\partial \ln C_l^{EE}}{\partial x_e(z_j)} = \sum_a \frac{ S_a(z_i) S_a(z_j)}{\sigma_a^2},
\eeq
where we have discretized the redshift space with $\delta z= 0.25$, and where $\sigma_a^2$ are the expected variances of the PCs.
The Fisher matrix is computed at a fiducial model which we take to be $x_e^{\rm fid}(z)=0.15$ in the parameterized range.
 Note that following Ref.~\cite{Heinrich:2018btc}, we have updated the PCs to be computed 
 with Planck 2015 rather than WMAP best-fit,
 which results in minor differences between our $S_a$ and those of  Ref.~\cite{Heinrich:2016ojb}.

We then rank-order the PCs from low to high variance, so that for the range $z_{\rm min} = 6$ (to be consistent with Ly$\alpha$ forest constraints, e.g.~\cite{Becker:2015lua}) and $z_{\rm max} = 30$, only the first 5 components are needed to describe all the information on $x_e$ carried by $C_\ell^{EE}$ to cosmic variance limit. In the work that follows, we therefore truncate the sum at $a = 1 .. N_{\rm{PC}}$, where $N_{\rm{PC}} = 5$ for $z_{\rm max} = 30$ in the fiducial analysis.\footnote{We also test the robustness of our fiducial analysis in \S \ref{sec:modelindependent} by taking 
 for $z_{\rm max} = 50$ where $N_{\rm{PC}} = 7$ is required at the cosmic variance limit.}

\begin{figure}
          \includegraphics[width=0.48\textwidth]{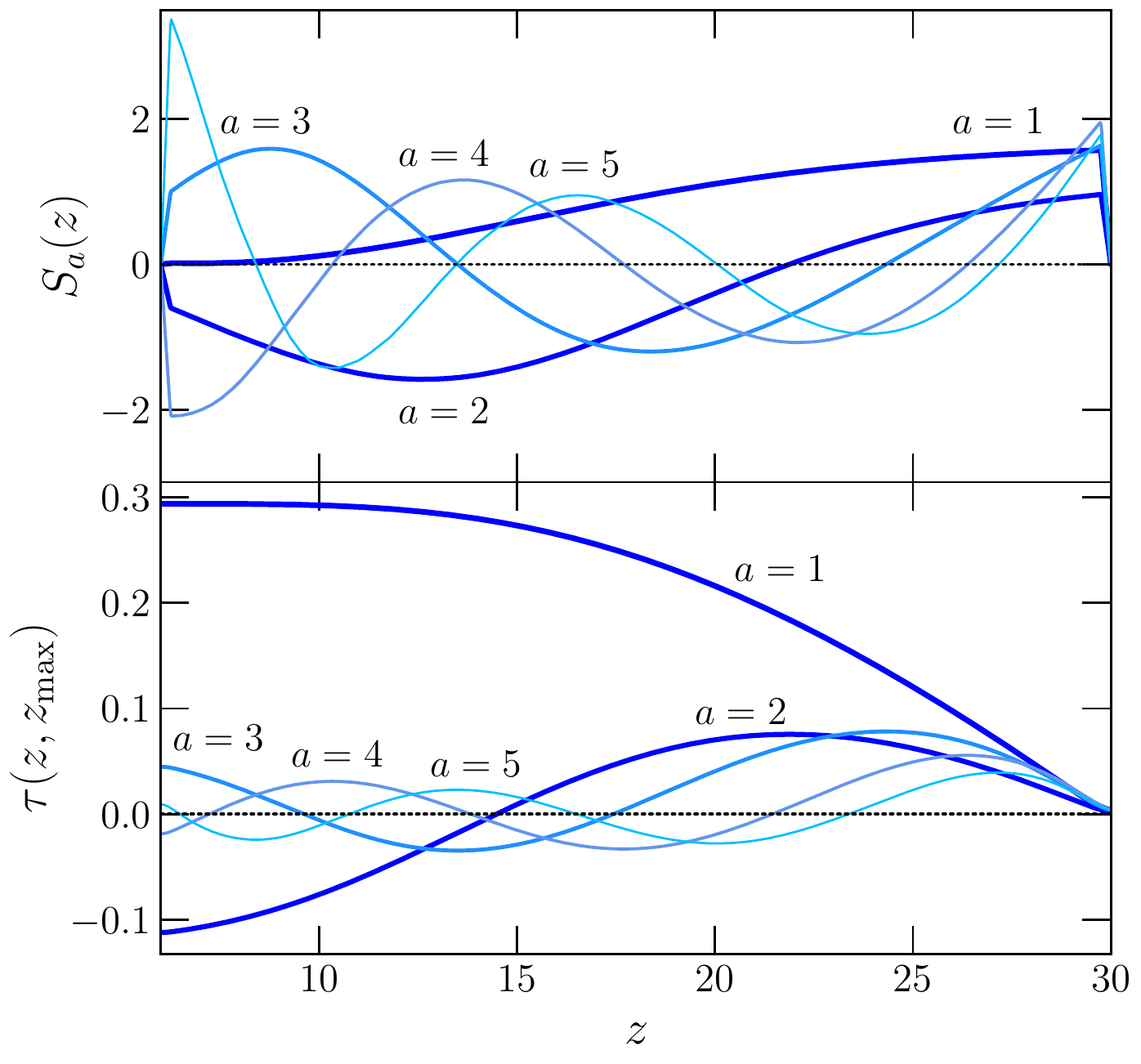}
          \caption{\textit{Top}: Rank-ordered CMB reionization principal components $S_a$ for $a = 1..5$ for $\zmax = 30$. We assume a fiducial model with fully ionized hydrogen and singly ionized helium by $z<6$, and doubly ionized helium at $z = 3.5$.
          \textit{Bottom}: The cumulative optical depth from $z$ to $z_{\rm max}$ of a unit amplitude PC. } 
          \label{fig:xe}
\end{figure}

 In Fig.~\ref{fig:xe} (top), we show the corresponding fiducial basis functions $S_a(z)$.  
Notice that the series resembles a Fourier decomposition of $x_e(z)$ rank ordered from low to high frequency.
This is because high frequency variations in $x_e$ leave little observable imprint in $C_l^{EE}$.  
As a consequence  the $N_{\rm PC}$ PCs are a complete representation of the \textit{observable impact} of $x_e(z)$ on the $C_\ell^{EE}$, rather than the ionization history itself. In other words, 
given any $x_e^{\rm true}(z)$, we can project it onto the $N_{\rm PC}$ PC basis through
\begin{equation}
m_{a}=
  \int _{\zmin}^{\zmax} dz\, \frac{S_{a}(z) [x_e^{\rm true}(z)-\xef(z)]}{\zmax-\zmin},
\label{eq:xetommu}
\end{equation}
where the reconstructed $x_e(z)$ through Eq.~(\ref{eq:mmutoxe}) with truncated PCs will not reproduce the true ionization history $x_e(z) \neq x_e^{\rm true}(z)$, but rather it is the observed $C_\ell^{EE}$ that is reproduced to cosmic variance precision. We reiterate therefore, that the PC analysis is not a tool for reconstructing the ionization history from observations, but rather a forward-modeling tool which, by reducing the dimensionality of the model space to $N_{\rm PC}$, allows us to constrain all possible ionization histories between $z_{\rm min}<z<z_{\rm max}$ in a single analysis.
For the Planck data set, most of the information on the ionization history comes from contraints on the first two modes which probe the amount of low vs.\ high redshift optical depth.   However all 5 PCs for $z_{\rm max}=30$ carry some constraint, with the higher modes controlling finer variations in redshift.

 We  compute the CMB power spectra with PCs using a modified version of CAMB\footnote{CAMB: \url{http://camb.info}}~\cite{Lewis:1999bs, Howlett:2012mh} (see \cite{Heinrich:2016ojb}).
 %\todo{CH: probably meant \cite{Heinrich:2016ojb}}.
 We follow CAMB and assume for $z<6$, fully ionized hydrogen and singly ionized helium whereas for $z\leq 3.5$~\cite{Becker:2010cu}, doubly ionized helium~\cite{Becker:2010cu} with a tanh transition width $\Delta z = 0.5$, given the helium fraction
 \beq
 f_{\rm He} = \frac{n_{\rm He}}{n_{H}} = \frac{m_{\rm H}}{m_{\rm He}} \frac{Y_p}{1 - Y_p}, 
 \eeq
as the ratio of the helium to hydrogen number density, where $Y_p$ is the helium mass fraction, chosen to be consistent with big bang nucleosynthesis for a given baryon density.

We obtain posterior constraints on the PC parameters $m_a$ from Markov Chain Monte Carlo sampling using CosmoMC\footnote{CosmoMC: \url{http://cosmologist.info/cosmomc}} \todo{CH:cite} of the relevant Planck likelihoods discussed in \S \ref{sec:results} in an otherwise fiducial $\Lambda$CDM cosmology with the standard 5 parameters: the baryon density $\Omega_b h^2$, cold dark matter density 
$\Omega_c h^2$, effective angular sound horizon $\theta_{\rm MC}$, scalar power spectrum log amplitude $\ln (10^{10} A_s)$ and
its tilt $n_s$.  We take flat uninformative
priors in $m_a$ but discuss additional priors imposed by
removing demonstrably  unphysical ionization histories in \S \ref{sec:cumulative}.

\subsection{Effective PC Likelihood}
\label{sec:KDE}
The completeness property of the PCs enables us to turn PC chains obtained from a MCMC run into an effective likelihood that can be used to test any reionization model with the CMB without a cumbersome reanalysis at the level of power spectra data. In the following, we briefly recap the kernel density estimate (KDE) technique used to build this likelihood, as well as a Gaussian approximation, and refer the readers to Ref.~\cite{Heinrich:2016ojb} for more details.

The PC chains are composed of $N_{\rm sample}$ samples of discrete values of $\mathbf{m}_i = \{m_1, \ldots, m_5\}$ along with multiplicities $w_i$ for $i = 1$...$N_{\rm sample}$. Given any physical ionization history $x_e(z)$, we first obtain its PC representation $\mathbf{m}$ using Eq.~(\ref{eq:xetommu}). Since $\mathbf{m}$ can take any continuous value, we approximate its effective likelihood with a kernel density estimate
\beq
{\cal L}_{\rm KDE}\left({\rm data}|\mathbf{m} \right)  = \sum_{i = 1}^{N_{\rm sample}} w_i K_f(\mathbf{m}-\mathbf{m}_i),
\eeq
where the overall normalization is arbitrary, and where we have chosen the smoothing kernel $K_f$ to be a  multivariate Gaussian with mean zero and covariance $f\mathbf{C}$, where $\mathbf{C}$ is the $N_{\rm PC} \times N_{\rm PC}$ covariance matrix estimated from the PC chains (see Table~\ref{tab:PC_stats}) and $f$ is a fraction smaller than 1.
For a Gaussian posterior, increasing the covariance by $1+f$ corresponds to increasing the standard deviation by approximately $1+f/2$. To minimize the amount of smoothing needed while still maintaining good accuracy in the tail of the distribution for any physical models, we oversample the PC distributions by running the PC chains far beyond convergence for about $N_{\sample} \sim 10^6$ chain samples. 
For this $N_{\sample}$, a fraction of $f = 0.14$ should be sufficient for most models, and indeed we have tested it to work well for the ones we study in \S~\ref{sec:effective_likelihood}.

An even simpler effective likelihood is the Gaussian approximation of the $m_a$ posterior using the PC mean $\bar{\bf m}$ and covariance $\bf C$
\begin{equation}
 {\cal L}_{\rm Gauss}\left({\rm data}|\mathbf{m} \right) = \frac{ e^{-\frac{1}{2} ({\bf m}-\bar{\bf m})^T {\bf C}^{-1} ({\bf m}-\bar{\bf m}) } }{\sqrt{(2\pi)^{N_{\rm PC}} | \mathbf{C}| }}.
 \label{eq:gaussian}
 \end{equation}

For any set of model parameters ${\bf p}$ with their given prior probability distribution
$P({\bf p})$, their posterior distribution can be approximated as 
\begin{equation}
P({\bf p}| {\rm data}) \propto {\cal L}_{\rm PC}\left[ {\rm data}|\mathbf{m}(\bf p) \right] P(\bf p),
\end{equation}
with PC\,=\,KDE or Gaussian.
This approach obviates the need to implement specific reionization models into CAMB and conduct separate CosmoMC sampling for each, thereby significantly reducing both human and computational effort.
In \S\ref{sec:effective_likelihood} we test the effective KDE and Gaussian likelihoods against an exact implementation
for two example models: the standard steplike model and a two-step model allowing for high-redshift ionization.

\subsection{Cumulative Optical Depth}
\label{sec:cumulative}

Although reionization PCs are mainly a tool for model testing using complete information from large angle CMB polarization, as they do not reconstruct the rapidly varying aspects of $x_e(z)$ itself, they do provide model-independent constraints on redshift integrated quantities such as  the cumulative Thomson optical depth
\begin{align}
\tau(z,z_{\rm max}) & = n_{\rm H}(0) \sigma_T \int_z^{z_{\rm max}} dz \frac{x_e(z) (1+z)^{2} }{H(z)},
\label{eq:cumtau}
\end{align}
where $n_{\rm H}(0)$ is the hydrogen number density at $z=0$, $\sigma_T$ is the Thomson scattering cross-section and $H(z)$ is the Hubble parameter. 
Given the tight constraints on cosmological parameters in the $\Lambda$CDM model, this quantity is well approximated by
\begin{align}
\tau_{\rm PC}(z,z_{\rm max})& = \sum_{a=1}^{N_{\rm PC}} m_a \tau_a(z,z_{\rm max}) + \tau_{\rm fid}(z,z_{\rm max}),
\label{eq:cumtauPC}
\end{align}
where $\tau_a$ and $\tau_{\rm fid}$ are defined by employing Eq.~({\ref{eq:mmutoxe}}) in
Eq.~(\ref{eq:cumtau}) using the fiducial model 
for $H(z)$ and number densities.  This differs from Ref.~\cite{Heinrich:2016ojb} where their cosmological dependence was retained in determining $\tau_{\rm PC}$.
 When we omit the redshift arguments, we
implicitly mean the total range, e.g. 
 $
 \tau_{\rm PC}\equiv \tau_{\rm PC}(0,z_{\rm max}).
 $
 
 In Fig.~\ref{fig:xe} (bottom), we display $\tau_a(z,z_{\rm max})$ for the $\zmax=30$ PCs.
 Notice that positive values of the first component $m_1$ mainly represents the optical depth that is accumulated at high redshift whereas {\it negative} values of $m_2$ provides the same for low redshift.    Since these two are the best measured components, the CMB mainly determines the amount of low vs.~high redshift ionization.
The underlying reason is that in $C_l^{EE}$ the higher the redshift, the
larger the relative contribution at higher $l$ due to the size of the horizon when the photons scattered.  For the less well measured PCs with $a>2$, the contributions to the total optical depth rapidly diminish and represent finer distinctions in redshift for the cumulative optical depth and multipole for the underlying $C_l^{EE}$ spectrum.

In the usual approach to constraining reionization in $\Lambda$CDM, one places an implicit prior on the shape of the cumulative optical depth by choosing a ``tanh" or near step function reionization (see \S \ref{sec:example1}) and then extracts a single constraint on the total optical depth.
The PCs, on the other hand, allow for arbitrary values of $x_e(z)$ when no range-bound prior constraints on the mode amplitudes $m_a$ are imposed. The one subtlety when placing constraints on the cumulative optical depth is that the analysis also allows
unphysical ionization fractions where $0 \le x_e \le x_e^{\rm max}$  is not satisfied.
Therefore, when determining cumulative optical depth constraints as opposed to using reionization PCs as a tool for testing models,
 we do impose a  prior
 by truncating the posteriors of $m_a$
 {\it after} obtaining them from Markov Chain Monte Carlo sampling, following Ref.~\cite{Mortonson:2008rx}
\begin{equation}
\sum_{a=1}^{N_{\rm PC}} m_a^2 \le (x_e^{\rm max}-x_e^{\rm fid})^2,
\label{eq:prior_sum}
\end{equation}
and 
%$x_e^{\rm fid}=0.15$ and 
$m_a^{-} \le m_a \le m_a^{+}$ with
\begin{equation}
m_a^{\pm} = \int_{z_{\rm min}}^{z_{\rm max} } dz \frac{S_a(z)[x_e^{\rm max} -2 x_e^{\rm fid}(z)]
\pm x_e^{\rm max} | S_a(z)|}{2(z_{\rm max}-z_{\rm min})}.
\label{eq:individualprior}
\end{equation}
Here we take $x_e^{\rm max} = 1+ f_{\rm He} $ to account for the contribution of singly ionized helium.
In the fiducial analysis, the prior on the sum of squares in Eq.~(\ref{eq:prior_sum}) is strictly weaker than the individual priors in Eq.~(\ref{eq:individualprior}).
Henceforth we refer to Eq.~(\ref{eq:individualprior}) as the physicality prior.  We study its impact and robustness in \S \ref{sec:modelindependent}.
Note that the original $0\le x_e \le x_e^{\rm max}$ condition cannot be strictly enforced here because we  keep only the first $N_{\rm PC}$ PCs, so these priors in $m_a$ are necessary but not sufficient conditions for a physical ionization history.  
In other words, no physical model is excluded by these priors but some unphysical models are included.  In \S \ref{sec:note_on_priors} we further explore the role of these and other choices of PC priors, and in \S\ref{sec:effective_likelihood} we test PC constraints against exact constraints for two example models.

\section{Planck 2018 PC Results}
\label{sec:results}
We now present the complete reionization constraints obtained from the Planck 2018 likelihoods using the principal components.
\subsection{Constraints on Principal Components}
We use the official Planck likelihoods~\cite{Aghanim:2019ame} \texttt{plik\_lite\_TTTEEE} for the high-$\ell$ $TT$, $TE$ and $EE$ as well as \texttt{lowl} for the low-$\ell$ $TT$ throughout this paper. We have tested that our results do not change if in lieu of \texttt{plik\_lite\_TTTEEE} we used the full \texttt{plik\_full\_TTTEEE} likelihood in which the foreground parameters have not been marginalized over. For the low-$\ell$ $EE$ likelihood, we use the third-party released \texttt{SRoll2} likelihood~\cite{Delouis:2019bub} in our official PC results and the effective likelihood code. In comparison with the official Planck-released \texttt{SimAll} likelihood, the \texttt{SRoll2} likelihood has improved foreground cleaning, 
which enables stronger reionization constraints, especially from the low multipole moments that are important for the low redshift constraints.

\begin{figure*}
\includegraphics[width=0.8\textwidth]{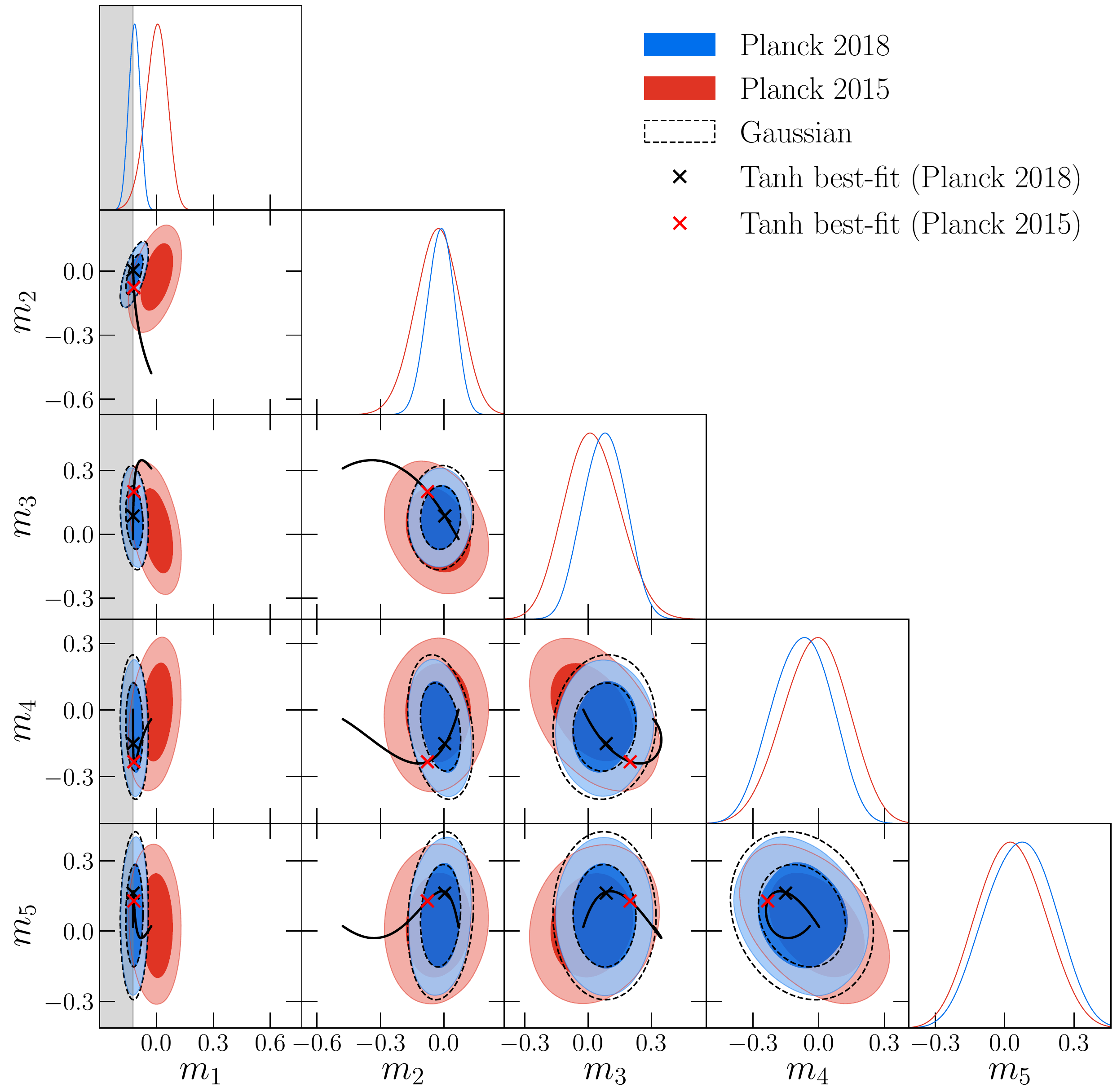}
\caption{Constraints from Planck data on the amplitudes of the five reionization principal components that describe all physical ionization models up to $z_{\rm max} = 30$. We show  the 68\% and 95\% $m_a-m_b$ constraints as well as marginal constraints on individual $m_a$ for Planck 2018 in blue,
using \texttt{SRoll2} for the low-$\ell$ $EE$ likelihood, and compare that to the Planck 2015 constraints in red.  The corresponding marginal distributions for individual $m_a$ are also shown.
When deriving model independent optical depth constraints we impose an additional physicality prior that excludes the gray region in $m_1$ and regions beyond the box limits in the others.
 The black solid line indicates the trajectory of tanh models with varying $\tau$. The red and black crosses indicate the best-fit tanh models from Planck 2015 and 2018 respectively. Note that the best-fit tanh model has moved to a lower $\tau$ in Planck 2018, and in particular, is now within the 68\% C.L. contours of the constraints.
}
\label{fig:plot_mjs_2018_vs_2015}
\end{figure*}
In Fig.~\ref{fig:plot_mjs_2018_vs_2015}, we show the 1D posterior and 2D 68\% and 95\% confidence level contours for the 5 PC amplitudes that describe ionization models up to $\zmax=30$. We also show the trajectory of the tanh model (see \S \ref{sec:example1}) through this space as well as the best fit tanh model points.\footnote{The Planck 2015 best-fit for the tanh model is derived from a chain best-fit model as taken from Ref.~\cite{Heinrich:2016ojb}, whereas for the Planck 2018 point in this paper we used a minimizer to find the best-fit; there are also slight model differences, where in Planck 2018 we used a tanh width of $\Delta z = 0.015(1+z)$ instead of $\Delta z = 0.5$ for the Planck 2015 best-fit. Using $\Delta z = 0.5$ for Planck 2018 moves the cross negligibly on this plot.}
The gray shaded region 
represents the parameter space that would be excluded by an additional 
physicality prior from Eq.~(\ref{eq:individualprior}),
which is just beyond the border of the displayed regime for all but $m_1$. Results using the Planck 2018 data are shown in blue. 

\begin{figure}
\includegraphics[width=0.4\textwidth]{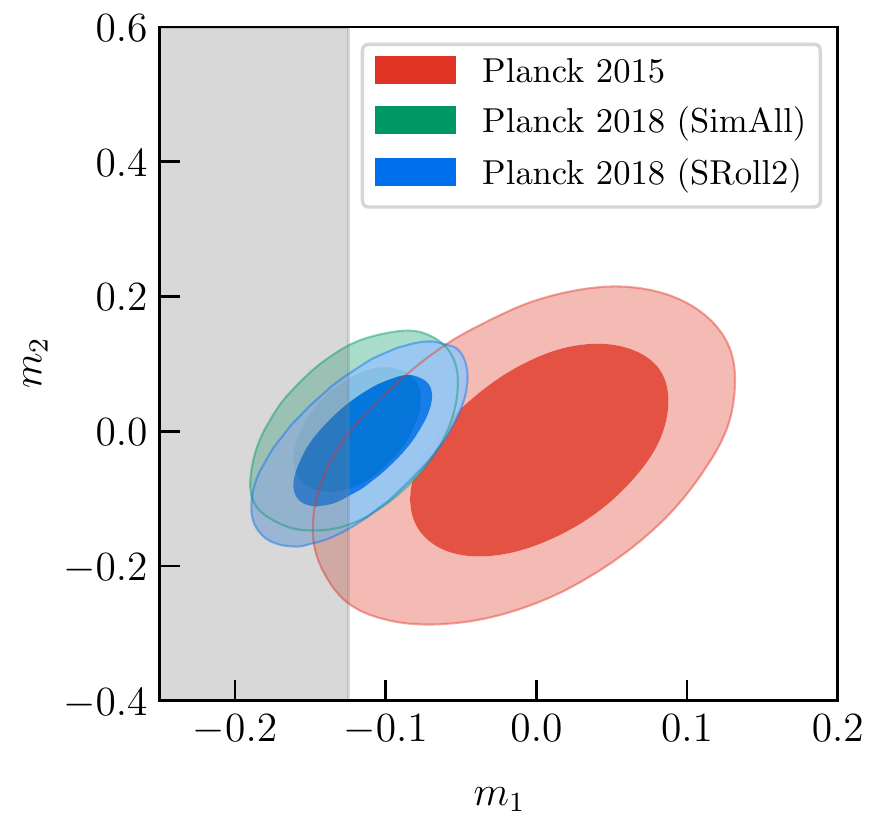}
\caption{Changes in the best constrained $m_1-m_2$ PC plane between Planck 2015 (red) and Planck 2018 (blue,  \texttt{SRoll2}; green, \texttt{SimAll}). The default \texttt{SRoll2} likelihood moderately improves the best constrained direction  and raises the optical depth for Planck 2018.}
\label{fig:plot_m1m2_2015_vs_2018}
\end{figure}

While all five PC amplitudes are  constrained by Planck, the first two are particularly well-constrained. Contrary to the Planck 2015 results of Ref.~\cite{Heinrich:2016ojb} shown in red, the 2018 data prefers a much smaller $m_1$, leading to a reduced optical depth, especially at high redshift.  
This preference also increases the impact of placing a physicality prior at low values of $m_1$ when deriving model-independent constraints as we shall discuss further in \S \ref{sec:modelindependent}.
We show a zoomed version of the $m_1-m_2$ plane in Fig.~\ref{fig:plot_m1m2_2015_vs_2018}.
In the physicality prior excluded region (gray), the ionization and the cumulative optical depth at high redshift becomes negative.

In Fig.~\ref{fig:plot_m1m2_2015_vs_2018}, we also show
the Planck 2018 constraints using the older \texttt{SimAll} likelihood.  The improvement 
from switching to the \texttt{SRoll2} likelihood mainly strengthens the upper bound in the best constrained direction in the $m_1-m_2$ plane which shifts constraints toward higher total optical depth (see also Fig.~\ref{fig:prior_box}).

%\wh{possibly move the discussion of trajectories into the model section}
%The trajectories shown in black lines are the steplike models of instantaneous reionization allowed by the Planck 2015 data with the arrow pointing in the direction of increasing optical depth. Clearly, these steplikes models are only a part of the allowed physical model space. By construction, they miss any high-redshift ionization component by assuming neglibile ionization levels before the transition redshift $z_{\rm re}$. 

In Fig.~\ref{fig:plot_mjs_2018_vs_2015}, we also show the constraints assuming a Gaussian posterior in $m_a$ using the PC chain covariance and means listed in Table~\ref{tab:PC_stats}.  The visual agreement with the full posteriors is remarkably good, especially in the well-constrained components.  This suggests that the simple Gaussian likelihood of Eq.~(\ref{eq:gaussian}) suffices for most constraints on models. We shall see in \S\ref{sec:effective_likelihood} that a quantitative comparison indeed shows good agreement between the KDE and Gaussian likelihoods, with only minor differences in the shape of their posterior inferences.

\begin{table}[H]
\centering
\caption{PC chain means $\bar m_a$, standard deviations $\sigma(m_a)$,  and correlation matrix $R_{ab}={C_{ab}/[\sigma(m_a)\sigma(m_b)]}$. $\tau_a$ is the optical depth contribution for $m_a=1$ 
which adds to that of the fiducial model, 
$\tau_{\rm fid}=0.08626$.
}

\label{tab:PC_stats}
\begin{tabular}{|r |r r r r r|}
\hline
		
			  &
\multicolumn{1}{|c}{$m_1$} & \multicolumn{1}{c}{$m_2$} & \multicolumn{1}{c}{$m_3$} & \multicolumn{1}{c}{$m_4$} & \multicolumn{1}{c|}{$m_5$} 
		\\ \hline

$m_1$ &
1.000 & 0.657 & $-$0.208 & $-$0.094 & 0.067 \\
$m_2$ & 
0.657 & 1.000 & 0.048 & $-$0.273 & 0.142 \\
$m_3$ &
$-$0.208 & 0.048 & 1.000 & 0.064 & $-$0.018 \\
$m_4$ & 
$-$0.094 & $-$0.273 & 0.064 & 1.000 & $-$0.195 \\
$m_5$ &
0.067 & 0.142 & $-$0.018 & $-$0.195 & 1.000 \\ \hline
$\bar m_a$ &
$-$0.116 & $-$0.016 & $$0.078 & $-$0.077 & 0.066  \\
$\sigma(m_a)$ & 0.030 & 0.063 & 0.098 & 0.131 & 0.145 \\
\hline
$\tau_{a}$ &
 0.29403 & $-$0.11227 & 0.04506 & $-$0.01898 & 0.00960 \\
\hline
\end{tabular}
\end{table}

\section{Model Testing with PCs}
\label{sec:effective_likelihood}

We now turn to constraining reionization models with PCs using the
effective likelihood approach described in~\S\ref{sec:effective_likelihood}.  The benefit of this approach is that
complete constraints on any model of the ionization history  between $6 \leq z \leq \zmax$ are obtained ahead of time and compressed into a simpler likelihood, so that when doing model testing one no longer needs to modify the $x_e(z)$ function and jointly sample the other cosmological parameters in for example CAMB, each time a new ionization model needs to be tested. These provide savings on human time in addition to the computing speed-ups from the likelihood evaluations.

Our code is available on GitHub at {\url{https://github.com/chenheinrich/RELIKE}}. The initial release corresponds to the fiducial results in this paper, i.e.\ derived from the Planck 2018 $\texttt{plik\_lite\_TTTEEE} + \texttt{lowl} + \texttt{SRoll2}$ likelihoods at $\zmax = 30$.  
This is also the default analysis for all results that follow except where noted.
The most important part of the code release is the python package called \relike\  which provides the PC parameters for a model and also implements the Gaussian effective likelihood. The user can specify a customized $x_e(z)$ function taking in reionization model parameters and use \relike\  to return the effective likelihood of the model. The returned likelihood is already marginalized over cosmological parameters. 
To obtain marginalized distributions more easily for high-dimensional parameter space, one may choose to sample \relike\ within a MCMC sampler such as Cobaya\footnote{Cobaya: \url{https://cobaya.readthedocs.io/en/latest/}}~\cite{Torrado:2020dgo,2020arXiv200505290T} or CosmoSIS\footnote{CosmoSIS: \url{https://bitbucket.org/joezuntz/cosmosis/wiki/Home}}~\cite{Zuntz:2014csq}.

We also provide a pre-assembled CosmoMC code using the CosmoMC's generic sampler, with which we have generated the results presented in this section. This CosmoMC code has an internal  implementation of both the Gaussian and KDE likelihoods in fortran, which is separate from our main python package that only supports the Gaussian likelihood. 

When using the KDE likelihood, we suggest using the default value of $f = 0.14$ to avoid over-smoothing parameter posteriors while maintaining accuracy during the KDE operation. All our KDE results in this section are computed with $f = 0.14$, which we shall soon show to work well in recovering results from an exact MCMC. Note that in both the Gaussian and the KDE modes, the model parameters or priors must be arranged to explicitly satisfy fully-ionized hydrogen and singly-ionized helium for $z\leq 6$. 

Next we demonstrate the use and the successful recovery of parameter posteriors using the effective likelihood code by comparing them to an exact MCMC analysis using the original Planck likelihoods for two examples: 1) The tanh model, which is the standard approach adopted in CAMB; 2) a two-step toy model which has one additional parameter allowing for a high-$z$ ionization plateau before the transition to full ionization in the canonical tanh model.

\subsection{Example 1: tanh model}
\label{sec:example1}

\begin{figure}
\includegraphics[width=0.48\textwidth]
{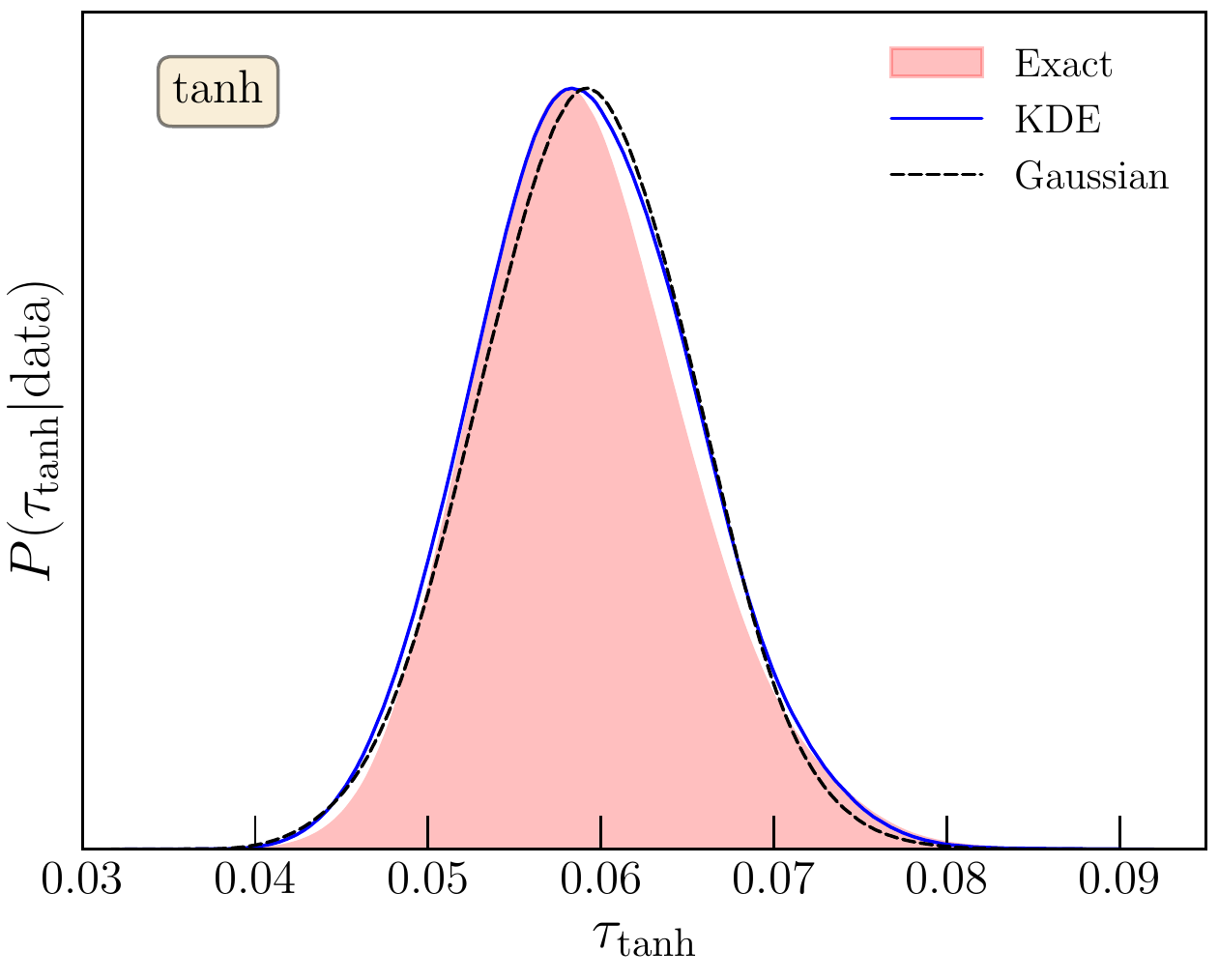}
\caption{Total optical depth $\tau_{\rm tanh}$ posterior for the tanh model in \refsec{example1} for the exact (Planck 2018) likelihood (shaded), KDE 
(blue solid) and  Gaussian (black dashed)
effective PC likelihoods. The overall agreement is excellent and the KDE results even capture the small skewness of the exact result.
}
\label{fig:tanh}

\end{figure}

\begin{figure}[ht]
\includegraphics[width=0.48\textwidth]{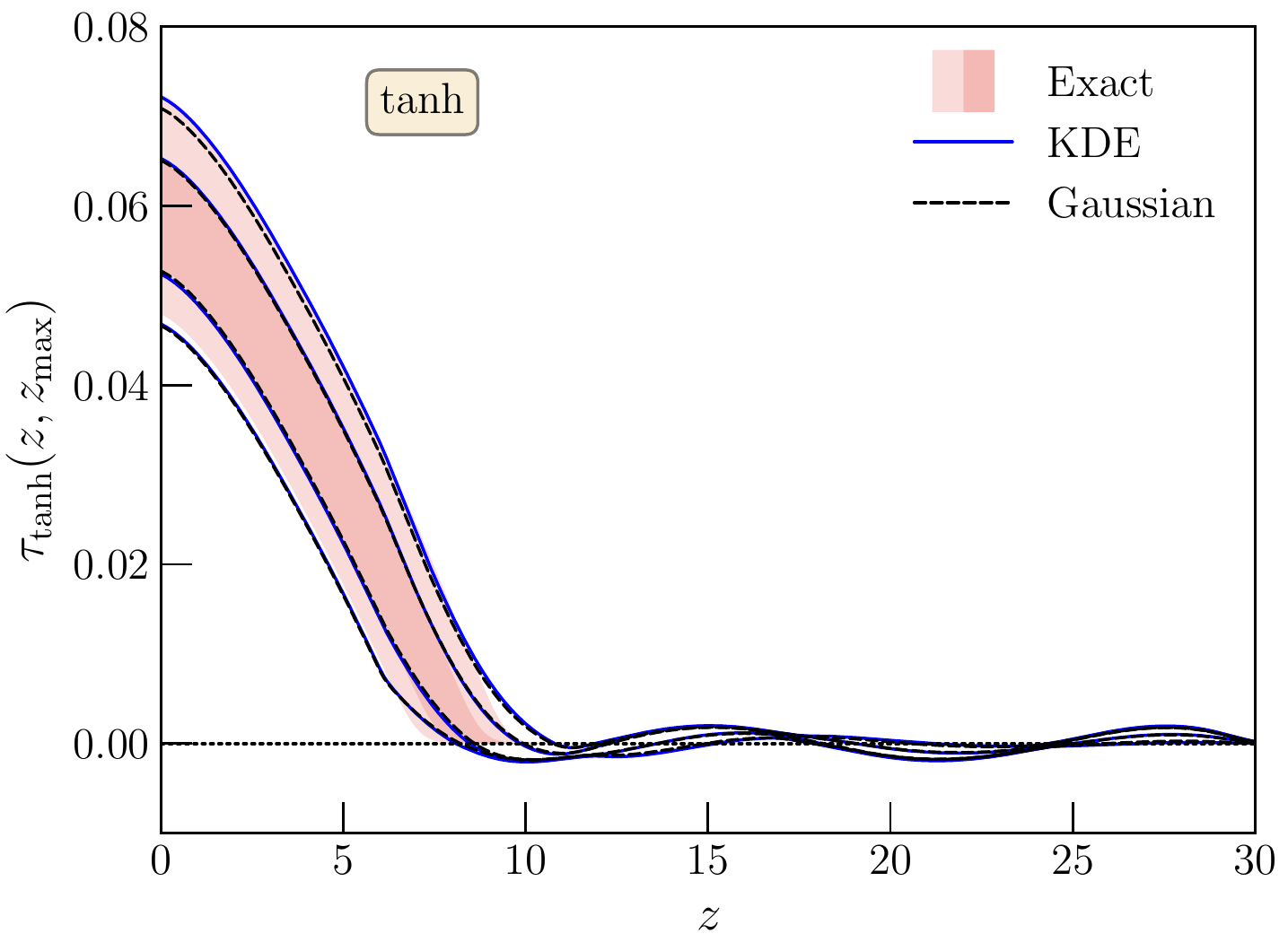}
\caption{Cumulative optical depth in the tanh model as in Fig.~\ref{fig:tanh}
(68\% and 95\% C.L. contours).  Overall agreement is good in regions where the contribution to $\tau_{\rm tanh}$ is high but the model form forbids high-$z$ contributions.    Oscillations around zero at high-z in the KDE and Gaussian PC representations are due to the effective low pass filter from the truncation at 5 PCs which does not produce any observable difference in the CMB. }
\label{fig:plot_taugtz_PC_vs_tanh}
\end{figure}

The steplike model used as the standard approach in CAMB describes the hydrogen and singly ionized helium reionization as a tanh function:
 \begin{equation}
x_e(z) = \frac{1+f_{\rm He}}{2}\left\{  1+ \tanh\left[ \frac{y(z_{\rm re})-y(z)}{\Delta y} \right] \right\} + x_e^{\rm rec},
 \label{eqn:tanh_1D}
 \end{equation}
 with $y(z)=(1+z)^{3/2}$, $\Delta y=(3/2)(1+z)^{1/2}\Delta z$. Instead of using $\Delta z = 0.5$ as in the standard setting by CAMB, we adopt a smaller width $\Delta z = 0.015(1+z)$ so as to better capture the required full ionization below $z=6$. Here $x_e^{\rm rec}$ is the ionization history from recombination only and we follow the fiducial prescription described in \S \ref{sec:background} for the 
 doubly ionized helium transition at $z=3.5$.

The $\tanh$ model is parameterized by its total optical depth for which we take a flat prior 
with an additional constraint that $z_{\rm re}\ge 6.1$ so as to  satisfy the  required full ionization by $z_{\rm min}=6$ given the tanh width to good approximation.

In Fig.~\ref{fig:plot_mjs_2018_vs_2015}, we show the trajectory that the $\tanh$ model takes in the PC space. Notice that this trajectory passes near the center of all constraints for Planck 2018 but only through the tails for some of the Planck 2015 constraints. Correspondingly, whereas the $\tanh$ model was marginally disfavored with Planck 2015 data, that is not the case with Planck 2018 data.

In \reffig{tanh} we show the posterior distribution of $\tau$ for the effective likelihood results, in which the only sampled parameter is $\tau$ (all other cosmological parameters are fixed at the Planck best-fit model), vs the exact Planck likelihood results in which the five other $\Lambda$CDM parameters were also varied using their standard priors. More specifically, we find
\begin{align}
    \tau_{\mathrm{tanh}} = 0.0591_{-0.0068}^{+0.0054}, & \quad (\mathrm{Exact})\nonumber \\
    \tau_{\rm tanh} = 0.0591_{-0.0066}^{+0.0061}, & \quad  (\mathrm{KDE}) \nonumber \\
    \tau_{\rm tanh} = 0.0592_{-0.0062}^{+0.0062}, & \quad  (\mathrm{Gaussian}),
\end{align}
where our exact result is also consistent with the tanh model constraint in Ref.~\cite{Pagano:2019tci} for the same combination of likelihoods.

The posteriors of all three likelihoods agree well. The KDE likelihood gives a slightly broader posterior because the kernel density estimate in the PC space introduces smoothing. We expect the smoothing to be roughly at the $\sim$7\% level in the parameter posterior corresponding to the smoothing factor of $1+f = 1.14$ we applied to the PC covariance during KDE. Indeed, the KDE result has a 4\% larger 68\% confidence region than the exact likelihood result. 

Instead of the KDE likelihood, we can alternatively use the Gaussian approximation given in Eq.~(\ref{eq:gaussian}). The Gaussian likelihood is faster to evaluate as it does not require looping over the entire $\sim 10^6$ points in the PC chain. 
The Gaussian likelihood result above gives symmetric tanh $\tau$ constraints where as the KDE result is able to capture some of the skewness in the exact distribution. The shift in the peaks seen in Fig.~\ref{fig:tanh} between Gaussian and the exact likelihoods, $\Delta \tau \sim 0.001$, is consistent with the level of asymmetric error bars in the latter. 

We also show in Fig.~\ref{fig:plot_taugtz_PC_vs_tanh} the cumulative optical depth obtained from directly integrating the tanh model in the exact likelihood chains vs that from the 5-PC representation of the tanh models in the effective likelihood chains using Eq.~(\ref{eq:cumtauPC}).
Notice that because of the truncation of the PC expansion at $N_{\rm PC}=5$,
this representation shows a small oscillatory cumulative $\tau$ at $z \gtrsim 10$ whereas the true model has negligible ionization there.   This is because the best constrained modes only represent the smooth features in the ionization history and truncation is similar to a low pass filter in Fourier space.  Though these zero mean, small amplitude oscillations do not bias results when averaged over redshift, this effect should be born in mind when interpreting the model-independent results in \S\ref{sec:modelindependent}.  

To quantify these residual oscillations we consider 
the 95\% upper limit on the high-redshift optical depth.
For both effective likelihoods  $\tau_{\rm tanh}(15, 30) < 0.002$  while $\tau_{\rm tanh}(15,30) \approx 0$
from the exact likelihood. 
By construction, the tanh model does not allow for a significant fraction of the total optical depth to come from high redshift.  As we shall see in the next example the data themselves do allow for a small but significantly larger fraction of the total optical depth to originate from high redshift, when the form of the model permits it.

\subsection{Example 2: the two-step model}
\label{sec:example2}

\begin{figure}
\includegraphics[width=0.48\textwidth]{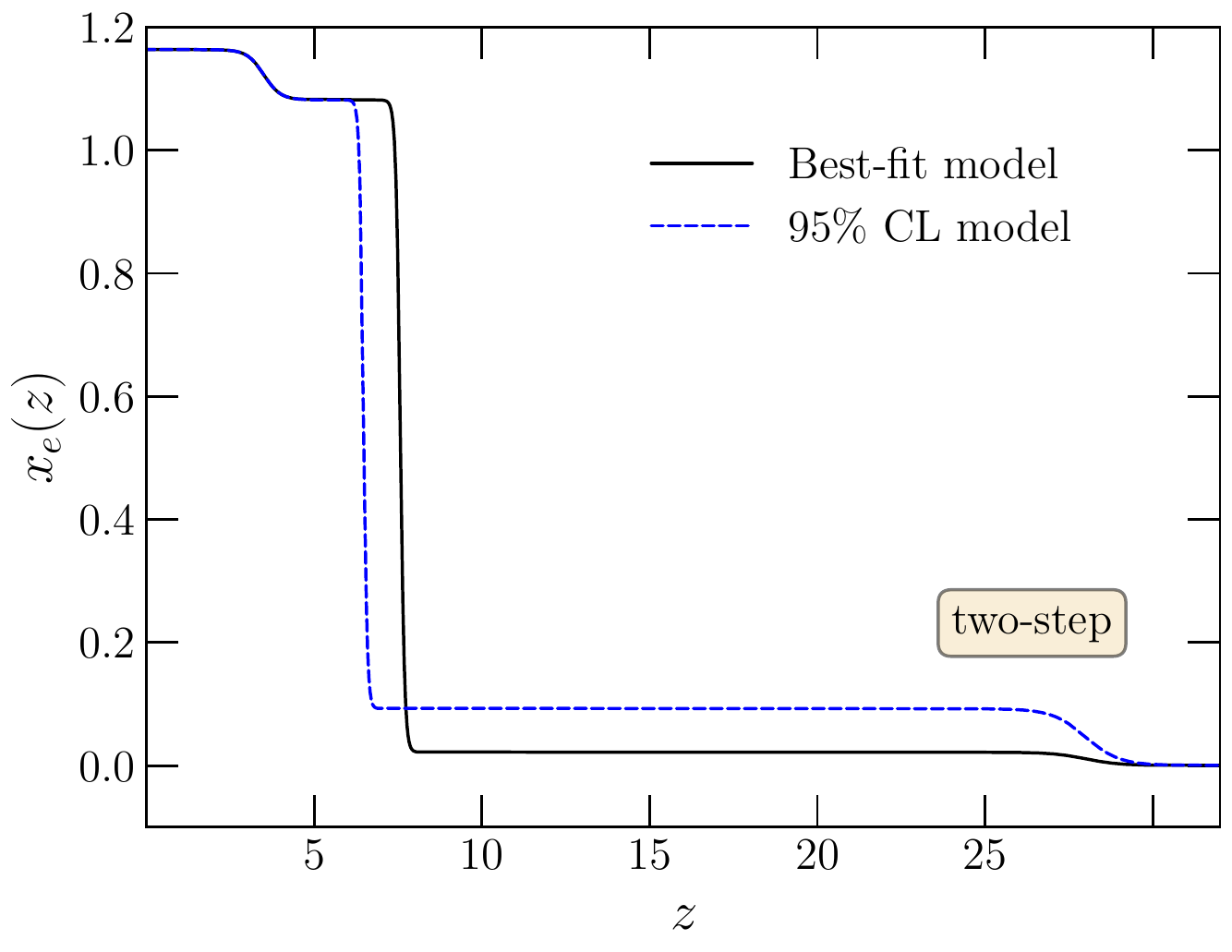}
\caption{Two-step ionization history $x_e(z)$ for the model of \refsec{example2}. In addition to a tanh transition at low-redshift as in \refsec{example1}, there is an additional ionization plateau at high redshifts with a fixed transition at $z_{\rm hi} = 28$ and $\Delta z_{\rm hi} = 1$. Here we show the best-fit model in the Planck 2018 data $(\taulo, \tauhi) = (0.053, 0.006)$, as well as  a model on the 95\% C.L. contour of Fig.~\ref{fig:two_parameter_model_2D} which maximizes $\tauhi$, $(\taulo, \tauhi) = (0.043, 0.026)$. }
\label{fig:two_step_model}
\end{figure}

\begin{figure}[t]
\includegraphics[width=0.48\textwidth]{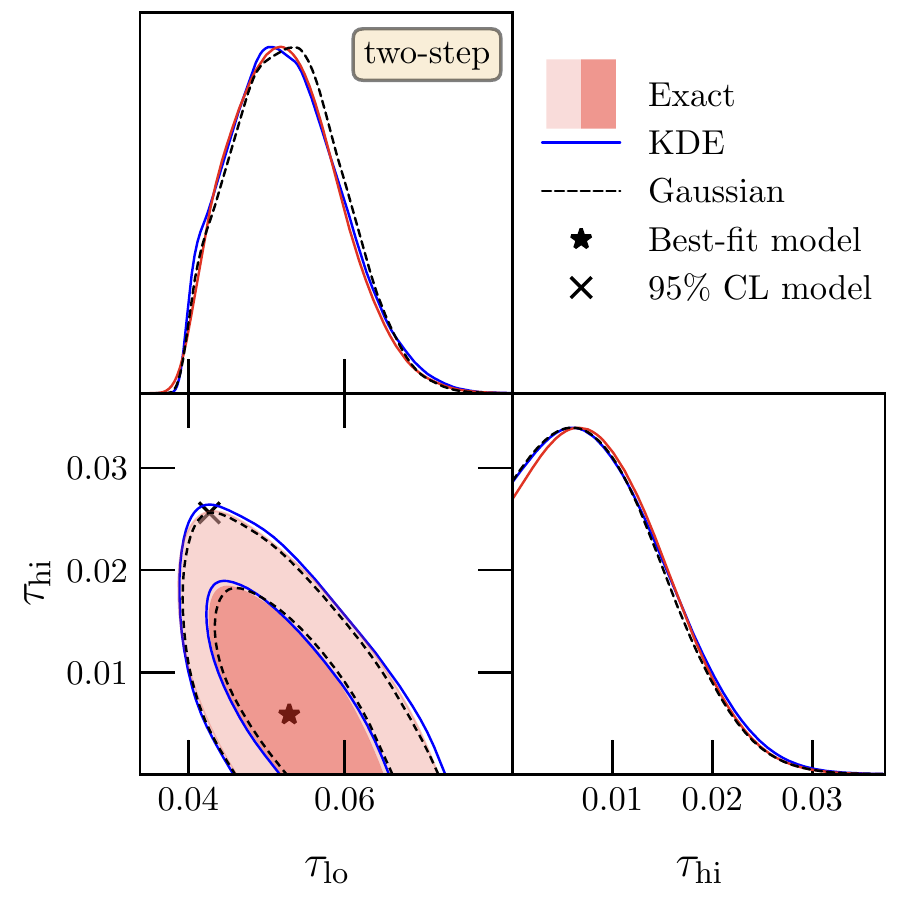}
\caption{Two-step parameter constraints (marginalized posteriors and 68\% and 95\% C.L. contours).  The exact, KDE PC and Gaussian PC likelihoods again agree very well and a small contribution from high-$z$ through $\tau_{\rm hi}$ is still allowed but not significantly favored.   Models of Fig.~\ref{fig:two_step_model} are marked and correspond to the best fit parameters and parameters 
which maximize $\tau_{\rm hi}$ on the 2D 95\% C.L. contour.
}
\label{fig:two_parameter_model_2D}
\end{figure}

\begin{figure}
\includegraphics[width=0.48\textwidth]{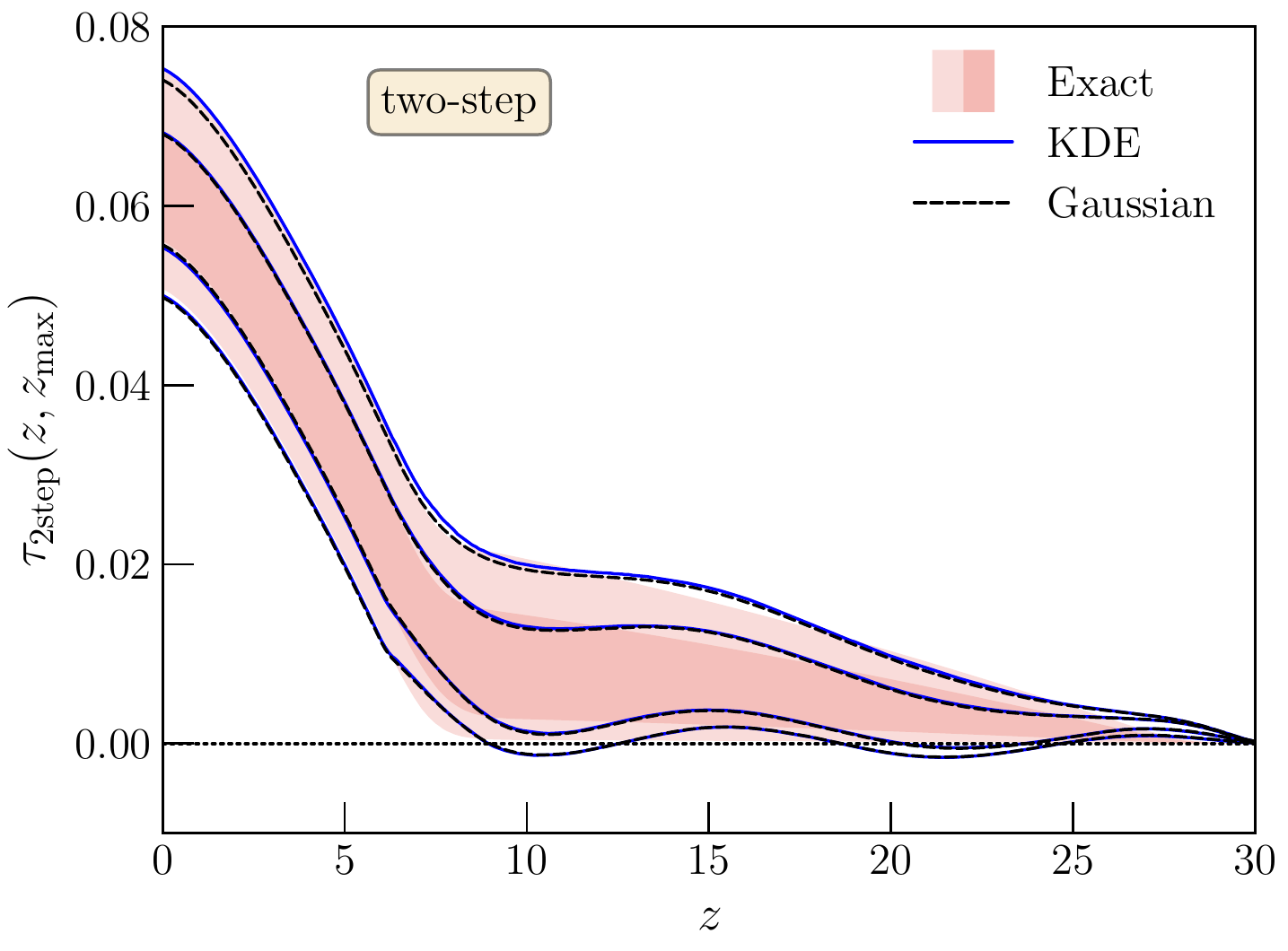}
\caption{Cumulative optical depth in the two-step model (as in Fig.~\ref{fig:plot_taugtz_PC_vs_tanh}).  Agreement between the exact, KDE PC and Gaussian PC likelihoods is again excellent and show that a small amount of high-$z$ ionization is allowed. The form of the model requires that the cumulative optical depth steadily declines above the first step at $z_{\rm re} \sim 6-7$ especially for $z\gtrsim 15$.
}
\label{fig:plot_taugtz_two_step_contours}
\end{figure}
 
The second example is a two-step model, where an additional step is used to give a high-$z$ ionization plateau: 
 \bea
x_e&(z)&\,= \frac{1+f_{\rm He} - \xemin}{2}\left\{  1+ \tanh\left[ \frac{y(z_{\rm re})-y(z)}{\Delta y} \right] \right\} \notag \\
&+& \frac{\xemin - x_e^{\rm rec}}{2}\left\{  1+ \tanh\left[ \frac{z_{\rm hi}-z}{\Delta z_{\rm hi}} \right] \right\} + x_e^{\rm rec},
 \label{eqn:tanh_highz}
 \eea
where $y(z)=(1+z)^{3/2}$, $\Delta y=(3/2)(1+z_{\rm re})^{1/2}\Delta z_{\rm re}$ where $\Delta z_{\rm re} = 0.015\, (1+z_{\rm re})$ just as in the first example for a sharper step.
We choose the second step at $z_{\rm hi}=28$ with $\Delta z_{\rm hi} = 1.0$, since the effective likelihood is limited to models with ionization below $\zmax=30$. 
In comparison to tanh, a single parameter $\xemin$ is added in this model to control the high-$z$ ionization plateau for $z_{\rm re} \lesssim z \lesssim z_{\rm hi}$. We would recover the standard tanh as $\xemin$ approaches the negligible recombination value $x_e^{\rm rec}$. We show examples
of this two-step model in Fig.~\ref{fig:two_step_model}.

In practice, we parameterize the model by $\taulo$ and $\tauhi$.  They are defined as in \refeq{cumtau} but with the boundaries $(0, z_{\rm split})$ and $(z_{\rm split}, \infty)$ for $\taulo$ and $\tauhi$ respectively, where the split is chosen as $z_{\rm split} = \zre + \Delta z_{\rm re}$ to be conservative on reducing the preference of $\tauhi$ in the data. 
Given a model $(\taulo, \tauhi)$, we find the corresponding ($\zre, \xemin$) through an iterative search similar to that used in the canonical tanh model for finding $\zre$ given a total optical depth $\tau$. 
For both the full likelihood and effective likelihood runs, we adopt the flat priors on $\taulo, \tauhi \in [0, 0.35]$.
We further apply a prior cut in $\zre$ space to keep models with $\zre > 6.1$ for both the full and effective likelihood chains as in the tanh model to ensure full reionization by $z = 6$ to good approximation. 

In ~\reffig{two_parameter_model_2D}, we display the posteriors of $\taulo$ and $\tauhi$. The exact, KDE and Gaussian PC treatments again agree very well. 
Note that in the canonical tanh model $\taulo \lesssim 0.04$ roughly corresponds to $\zre < 6$ so that the posteriors are effectively cut off  near this value.
The models in Fig.~\ref{fig:two_step_model} are represented by the star and cross symbols for the best fit case and a model allowed at the 95\% C.L. respectively.

The marginalized 1D constraints from the exact, KDE and Gaussian likelihoods also agree well
\begin{align}
\taulo = 0.0526^{+0.0058}_{-0.0078}, \; \tauhi < 0.021\; (95\% \text{ C.L.}), & \quad({\rm Exact}) \nonumber\\
\taulo = 0.0525^{+0.0062}_{-0.0079}, \; \tauhi < 0.021\;  (95\% \text{ C.L.}), & \quad({\rm KDE}) \nonumber\\
\taulo = 0.0528^{+0.0064}_{- 0.0077}, \; \tauhi < 0.021\; (95\% \text{ C.L.}), & \quad({\rm Gaussian}).
\end{align}
Note that the upper limit on $\tau_{\rm hi}$ presented here is the one-sided 95\% C.L. limit in 1D; whereas the model shown in Figs.~\ref{fig:two_step_model} and~\ref{fig:two_parameter_model_2D} is located on the 95\% C.L. contour in 2D and has therefore a larger $\tau_{\rm hi} = 0.026$.

Moreover, we can derive the constraints on the total optical depth, which are also consistent between the three likelihoods as well:
\begin{align}
\tau_{\rm 2step} = 0.0621^{+0.0062}_{-0.0061}, & \quad (\text{Exact)} \nonumber\\
\tau_{\rm 2step} = 0.0619^{+0.0063}_{-0.0065},  & \quad (\text{KDE)} \nonumber\\
\tau_{\rm 2step} = 0.0618^{+0.0062}_{-0.0062}, & \quad (\text{Gaussian).}
\end{align}
Given the small amount of $\tau_{\rm hi}$ still allowed, the total optical depth constraints in the two-step model are correspondingly slightly larger than $\tau_{\rm tanh}$.
This is consistent with other examples in the literature where extended reionization to high redshifts were allowed (e.g. Refs.~\cite{Ahn:2020btj, Paoletti:2020ndu}). More generally,
models which are sufficiently flexible at allowing for high-redshift ionization will have a comparable difference in their total optical depth.

Finally the 95\% C.L. upper limit on high-redshift optical depth are also consistent:
\begin{align}
\tau_{2\text{step}}(15, 30) < 0.014\; (95\% \text{ C.L.}), & \quad (\text{Exact}) \nonumber\\
\tau_{2\text{step}}(15, 30) < 0.016\; (95\% \text{ C.L.}), & \quad (\text{KDE}) \nonumber\\
\tau_{2\text{step}}(15, 30) < 0.016\; (95\% \text{ C.L.}), & \quad (\text{Gaussian}).
\label{eq:2stepbound}
\end{align}
The small difference in PC results vs.~exact reflects the mild oscillatory PC artifact around $z\sim 15$ in Fig.~\ref{fig:plot_taugtz_two_step_contours}, where we show the cumulative optical depth constraints on this two-step model in the exact vs the PC likelihood construction. Note that the 95\% upper limit on $\tau(15, 30)$ reported in the text are one-sided upper limits, whereas the contours in the cumulative optical depth figures enclose their respective C.L.
Note also that the 95\% C.L. allowed two-step model in 
Fig.~\ref{fig:two_step_model} has
$\tau_{2\text{step}}(15,30)= 0.017$.

The Planck 2018 cosmological parameter paper~\cite{Aghanim:2018eyx} also constrained $\tau(15, 30)$ in a model-independent way using the FlexKnot method, in which $x_e(z_i)$ at the $i$-th knots are varied as free parameters as well as the number of total knots. They obtained $\tau(15, 30) < 0.007$ (95\% C.L.), which would strongly rule out models that are still allowed at the 95\% C.L. in this class. 
In principle, the FlexKnot method allows for all possible physical ionization histories that increase monotonically with redshifts, and would include the two-parameter model described here. But apparently the implicit prior employed by the method disfavors optical depth at high redshift more than a flat prior in 
$\tau_{\rm hi}$ does.  We shall see in \S \ref{sec:modelindependent} that our model independent PC approach models such a flat prior better and of course the effective likelihood approach allows for any desired prior.

\begin{figure}[ht]
\includegraphics[width=0.48\textwidth]
{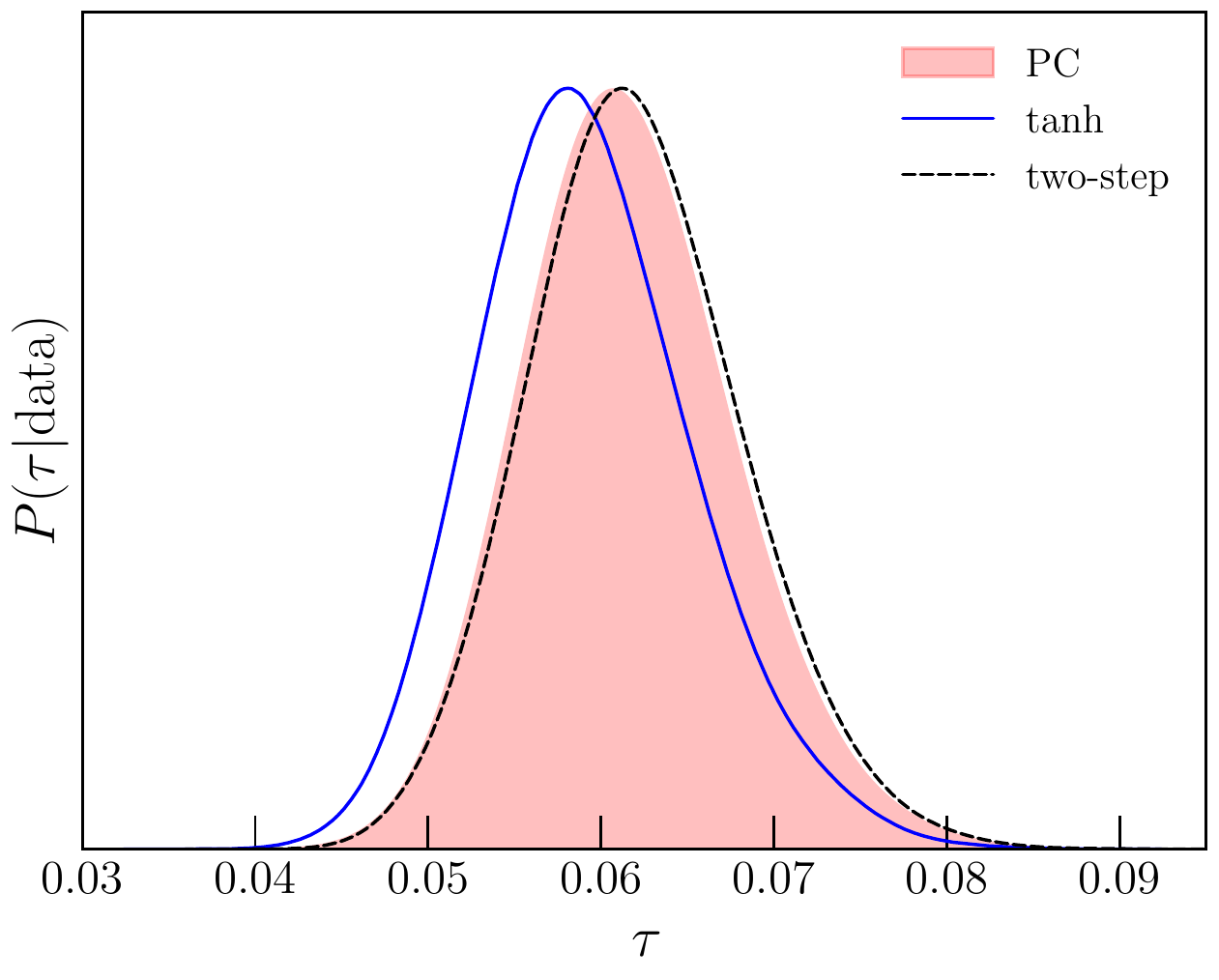}
\caption{Total optical depth posterior for the model-independent PC analysis (shaded) compared with the exact tanh model (blue solid line) and the two-step model (black dashed line) from Figs.~\ref{fig:tanh} and~\ref{fig:two_parameter_model_2D}. The PC analysis captures the ability to raise the optical depth through the high-$z$ contributions in models whose form allow it like in the two-step case.
}
\label{fig:tauPC}
\end{figure}

\section{Model-Independent Results for Optical Depth}
\label{sec:modelindependent}

While the PC approach to constraining reionization should mainly be used for 
testing wide classes of models efficiently and with their own physically motivated priors, it can also be employed to extract model-independent constraints on the optical depth. 
In this section we present these model-independent results, 
first on the total optical depth and then on the cumulative optical depth at high redshift, and discuss their robustness to priors  on physicality of the ionization history and redshift range of the fiducial analysis.

\subsection{Total optical depth}

\label{sec:note_on_priors}

Using Eq.~(\ref{eq:cumtauPC}), we can derive constraints on the total optical
depth $\tau_{\rm PC}=\tau_{\rm PC}(0,z_{\rm max})$ with the additional physicality priors on the PC amplitudes $m_a$ from
Eq.~\ref{eq:individualprior}.  These 
correspond to 
\beq
\tau_{\rm PC} = 0.0619^{+0.0056}_{-0.0068},
\label{eq:tau_pc_zmax30}
\eeq  
and the full posterior is shown in Fig.~\ref{fig:tauPC}. 
Notice that the distribution matches that of the two-step model from \S \ref{sec:example2} but almost $0.5\sigma$ shifted higher than that of the tanh model from \S \ref{sec:example1}.

The similarity and difference between these results come down to implicit and explicit priors. As we have seen, 
results based on the tanh model requires that reionization happens suddenly at $z_{\rm re}$ and is maintained thereafter.   Therefore, the model has a strong implicit prior that a given a total optical depth $\tau$ comes from $z\lesssim z_{\rm re}$ regardless of whether the data allow or prefer
ionization at higher $z$. While this prior may be well motivated theoretically, it should be distinguished from constraints that are data driven.  
This was especially apparent with the Planck 2015 data where the PC constraints gave $\tau_{\rm PC}=0.092\pm0.015$ which differed substantially from the tanh $\tau_{\rm PC}=0.079\pm0.017$~\cite{Heinrich:2016ojb}.  
As we shall discuss below, these differences were even more striking in the implications for the cumulative $\tau$ at high redshift as we have also concretely illustrated in the two-step model of \S \ref{sec:example2}.

PC constraints on the total optical depth do not suffer from assuming a specific class of ionization histories as they parameterize any $x_e(z)$ from $z_{\rm min}=6$ to $z_{\rm max}=30$.
They do of course come with their own priors namely the flat
range-bound ones from Eq.~\ref{eq:individualprior}.
These have in fact caused some confusion in the literature when interpreted in terms of the total optical depth \cite{Millea:2018bko}.  Here we first review the impact of flatness in PC space as discussed extensively in Ref.~\cite{Heinrich:2018btc} and then highlight the more important role of physicality priors with Planck 2018 data as compared with
2015 data.

\begin{figure}[ht]
\includegraphics[width=0.48\textwidth]{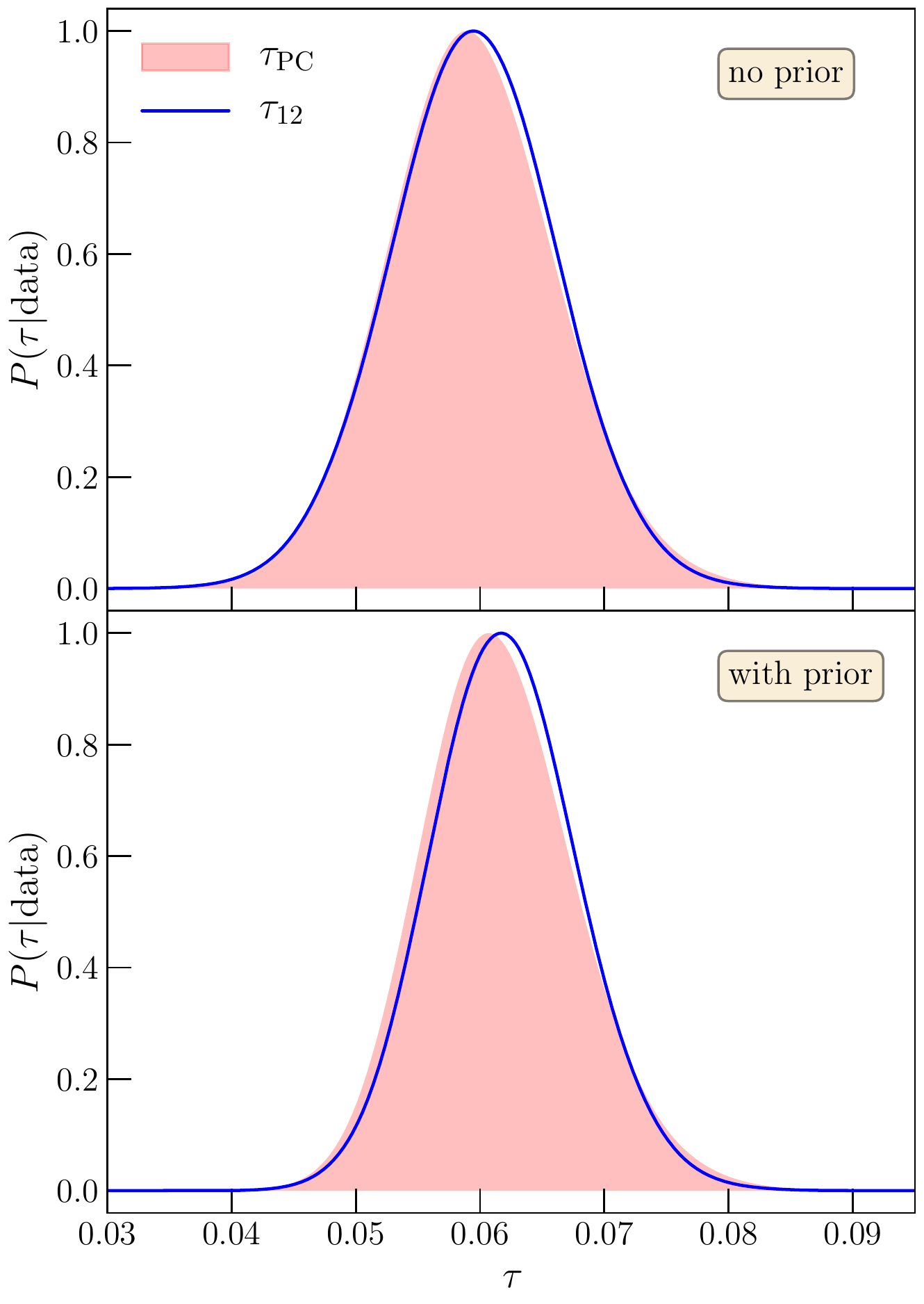}
\caption{Total optical depth $\tau_{\rm PC}$ posterior (shaded) without (top) and with (bottom) the additional physicality prior of Eq.~(\ref{eq:individualprior}).  
Imposing the physicality prior eliminates negative ionization at high $z$ and shifts the distribution to a higher total $\tau$. This unphysical contribution comes from the $m_1$ PC, as can be traced through the approximate $\tau_{12} \approx \tau_{\rm PC}$ built from a linear combination of the first two components defined in Eq.~(\ref{eq:tau12}) (blue lines).}
\label{fig:tau12}
\end{figure}

 We first note that by virtue of Eq.~(\ref{eq:cumtauPC}), the total optical depth receives contribution from all PCs.  However the PCs are rank ordered by their expected constraining power and so most of the information on the total optical depth comes from
the first two components
\beq
\tau_{\rm PC} \approx \tau_{12} \equiv  m_1 \tau_1 + m_2 \tau_2 +
\left(\sum_{a=3}^5 \bar m_a \tau_a + \tau_{\rm fid}\right) ,
\label{eq:tau12}
\eeq
where the term in parentheses is a fixed quantity (see Tab.~\ref{tab:PC_stats}) whose purpose is to remove offsets due to the arbitrary choice of the fiducial ionization model.  We explicitly demonstrate this approximate equivalence in Fig.~\ref{fig:tau12} by comparing the
posterior distributions of $\tau_{\rm PC}$ and $\tau_{12}$ both with and without the physicality prior.

Due to this linearity, flat priors in $m_1$ and $m_2$ correspond to
flat priors in $\tau_{12}$ for any linear trajectory through the space.   For example, the tanh model trajectory is shown in Fig.~{\ref{fig:plot_mjs_2018_vs_2015} and is nearly linear through the Planck 2018 allowed region in the $m_1-m_2$ plane.  Even without the explicit 
construction of a flat tanh $\tau$ prior in \S \ref{sec:example1}, 
the implicit prior from $m_1,m_2$ is already nearly flat. 

Ref.~\cite{Millea:2018bko} raised a potential objection regarding more general models.  As illustrated in Fig.~\ref{fig:prior_box}, lines of constant $\tau_{12}$ are rotated compared with the $m_1$,
$m_2$ axis.  Therefore, the range-bound flat prior in $m_1$, $m_2$ (blue square)  is not flat in its {\it marginal} $\tau_{12}$  distribution due to the extent  of the allowed parameter space along lines of constant $\tau_{12}$: At 
the implied lowest $\tau_{12}=0.023$ of the range
in Fig.~\ref{fig:prior_box},
this extent vanishes and then increases linearly until $\tau_{12} = 0.130$.
 Ref.~\cite{Millea:2018bko} suggests 
inverting this prior point-by-point in the $m_a$ parameter space by dividing by the implied prior $P(\tau_{\rm PC}) [ \approx P(\tau_{12})]$; this reweighting was implemented in the Planck 2018 paper on cosmological parameters~\cite{Aghanim:2018eyx}.
First, notice that this inversion would 
produce a non-flat prior on $\tau$ for the tanh model
that disfavors high values of $\tau$.   This is because
the additional parameter space allowed by the range-bound $m_a$ prior is excluded by the functional form of the model, even though the reweighting occurs globally and includes the tanh trajectory as well. 

More generally, to the extent that the data constrain $m_1$ and $m_2$ better than the flat range-bound priors, these priors become irrelevant and the reweighting of Ref.~\cite{Millea:2018bko} becomes erroneously motivated.   To see this, consider a range-bound prior that is rotated to be oriented along constant $\tau_{12}$ (see Fig.~\ref{fig:prior_box}, blue dashed rectangle). This rotated prior in $m_1-m_2$ space corresponds to a constant $P(\tau_{12}$) once the direction orthogonal to $\tau_{12}$ is marginalized.
Had the data constraint been shifted to slightly higher
$m_1$ so that both prior ranges encompassed all of the 95\% C.L. allowed
region they both would yield  the same constraints at that level.
However the reweighting by $1/P(\tau_{12})$ would change the original prior but not the rotated one and destroy this desired equivalence. 
More generally the point is that for a multidimensional space, constructing a flat marginal 1D distribution per parameter in the absence of data does not guarantee a locally flat prior in the region supported by the data
(see \cite{Heinrich:2018btc} for further discussion).

The real impact of our range-bound $m_1,m_2$ priors comes from the fact that with the Planck 2018 data, the physicality prior on $m_1$
removes nearly half of the space that would otherwise be allowed by the
data at 95\% C.L.  This is in contrast to the Planck 2015 data where only a very small region is removed (see Fig.~\ref{fig:plot_m1m2_2015_vs_2018}).  
This is reflected in the upward shift of the $\tau_{12}$ posterior in Fig.~\ref{fig:tau12} with and without the physicality prior on $m_1$.   This prior excludes very negative $m_1$ and we can see from Fig.~\ref{fig:xe} that these cases would have an unphysically negative ionization and cumulative optical depth at high $z$.  More precisely, even though a very positive value 
of $m_2$ can restore physicality at the highest redshifts, the fact that there is a node near $\tau_2(15,z_{\rm max})$ means that no choice can counteract $m_1$ there.  
Correspondingly, the impact of the $m_1$ prior is to remove
cases where the high redshift contribution is negative which would otherwise broaden and shift the $\tau_{12}$ posterior to lower values.

This also shows how our range-bound priors 
are conservative: all the cases that are excluded are definitely unphysical but some of the cases that are included
may also be unphysical since their physicality depends on higher-order PC components.  The impact on optical depth  however is small and at most, contributes a small bias that conservatively {\it lowers} the high redshift contribution which is already a small contribution to the total.  
This can be seen  in the comparison with the two-step model 
in Fig.~\ref{fig:tauPC} and as we shall return to this point in the cumulative optical depth constraints in the next section where the high and low redshift contributions are manifest.
In any case, using PCs to 
test explicitly physical models as in \S \ref{sec:example1} and
\ref{sec:example2} cannot be affected by our conservative physicality
cut.

Finally, we check that our results (Eq.~\ref{eq:tau_pc_zmax30}) are robust to
 extending the PC
range to
$z_{\rm max}=50$ which is negligibly different
\beq
\tau_{\rm PC} = 0.0626^{+0.0061}_{-0.0072}\quad (\zmax=50).
\eeq
Reverting the Planck 2018 likelihood to \texttt{SimAll} at $z_{\rm max}=30$ would give
\beq
\tau_{\rm PC} = 0.0582 ^{+0.0072}_{-0.0083}\;\; (\texttt{SimAll}).
\eeq
As with the tanh model, the \texttt{SRoll2} likelihood provides tighter constraints that shift the distribution toward higher optical depth.

 \begin{figure}
          \includegraphics[width=0.9\columnwidth]{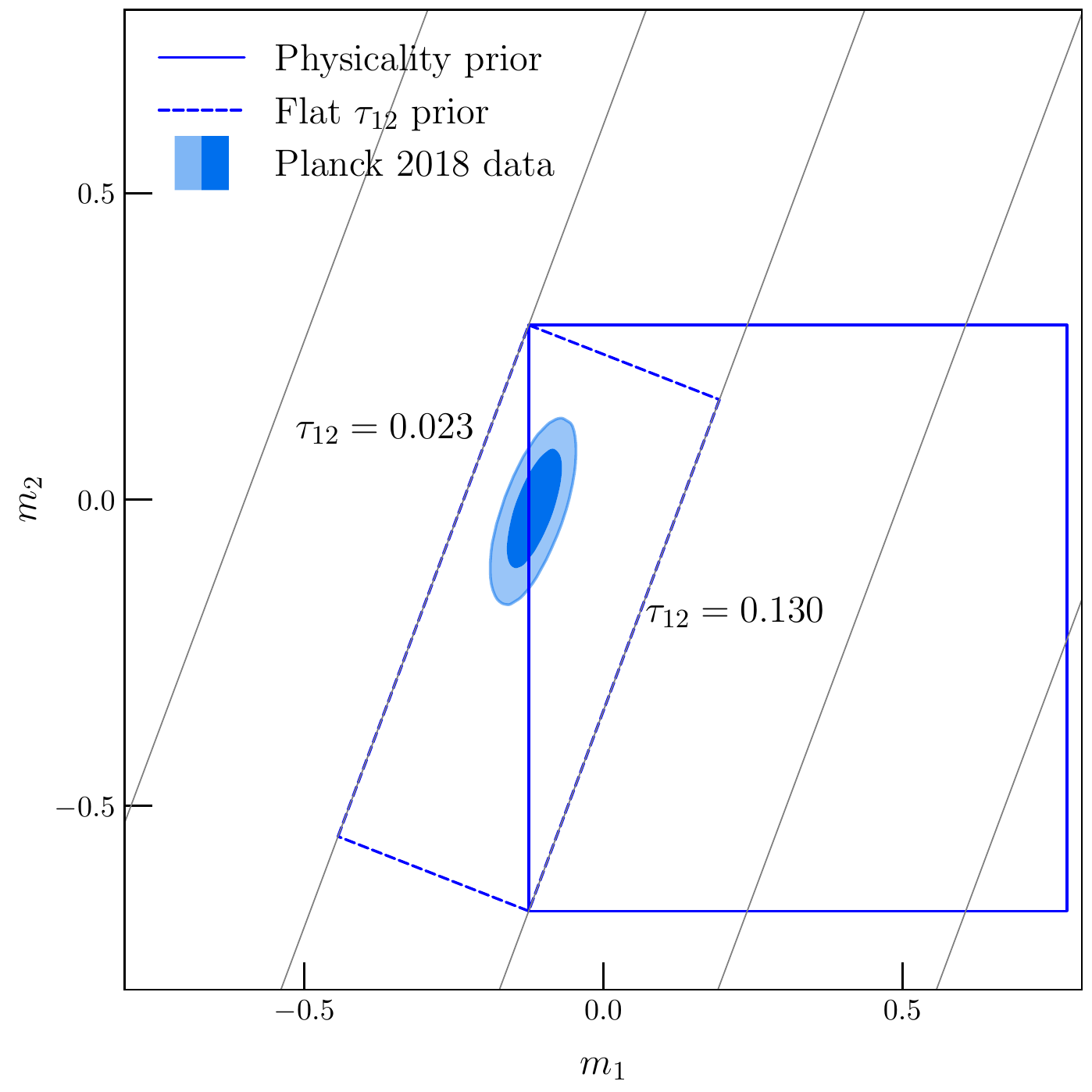}
          \caption {Priors on the $m_1-m_2$ parameter space: the default physicality prior (inset square, solid lines) and an alternate prior which is  flat in its marginal
           $\tau_{12}$ distribution (rotated rectangle, dashed lines) given boundaries
           which are aligned with lines of constant $\tau_{12}$ (light gray lines). Note that although the physicality prior allows for more parameter space at higher $\tau_{12}$, that is also the region excluded by the data constraints (ellipses). The only difference in the prior across the allowed region is that the physicality prior removes cases where $m_1$ is so small that it requires negative ionization at high-$z$.} 
          \label{fig:prior_box}
\end{figure}

\subsection{Cumulative optical depth}

The value of a model-independent approach is even more apparent for constraints on the
cumulative optical depth $\tau(z, \zmax)$ at high $z$.
The main results are again related to the $m_1$ and $m_2$ PC constraints, both of which are well constrained.  
Consequently, as shown in Fig.~\ref{fig:prior_box}, beyond the well-constrained total optical depth (as represented by the $\tau_{12}$ linear combination in Eq.~(\ref{eq:tau12})) 
is a somewhat weaker constraint in the constant $\tau_{12}$ direction that reflects whether that $\tau_{12}$ comes mainly from high or low redshift (see \S \ref{sec:cumulative}).
Any finer redshift detail in the ionization history is much more poorly constrained.

In Fig.~\ref{fig:plot_taugtz_2018_with_vs_without_physicality_prior}, we show the 
the 68\% and 95\% confidence level contours for  $\tau_{\rm PC}(z,z_{\rm max})$ with and without the physicality prior on $m_1$.  Notice that without the prior, a larger range of unphysically negative optical depth is allowed, whereas with the prior it is mostly eliminated except for the highest redshifts.  Moreover, upper bounds on the optical depth at high 
redshift are largely unchanged.   These upper limits allow some contribution at high-$z$ but do not require it.   For example,
\beq
\tau_{\rm PC}(15, 30) < 0.020 \; (95\%\; \mathrm{C.L.}).
\label{eq:tauPC15}
\eeq
Recall again that we are reporting the one-sided upper limit here in the text, which is slightly lower than values at the contours shown in the figures.
This should be contrasted with the tanh model 
where  high-$z$ ionization is forbidden by the functional form of the ionization history.
On the other hand the bound is comparable to that obtained for the two-step model (see Eq.~(\ref{eq:2stepbound})).  It is actually slightly larger given that the form of the two-step model still requires a steadily decreasing cumulative optical depth at $z>z_{\rm re}$ with the additional allowed contributions reflected in $\tau_{\rm hi}$.
 
These results are in contrast to those of the Planck 2015 analysis. In Fig.~\ref{fig:plot_taugtz_2015_vs_2018_simallEE_vs_2018_srollv2} top panel, we compare
the two. Whereas the Planck 2018 has only upper limits on the high redshift optical depth, Planck 2015 actually preferred
finite optical depth at $z\gtrsim 15$.

In the lower panel, we compare the Planck 2018 results for the two different low-$\ell$ $EE$ likelihoods.  Improvements from \texttt{SRoll2} come mainly from improving the lower bound at low redshifts, consistent with the improved constraints at the lowest multipoles of $C_l^{EE}$.

We also show using a separate PC chains with $\zmax = 50$ that there is no evidence for optical depth at $z>30$. 
In Fig.~\ref{fig:plot_taugtz_zmax30_vs_zmax50}, we show that the 68\% and 95\% C.L. contours of $\tau(z, \zmax)$ are consistent between the $\zmax = 30$ (red regions) and $\zmax = 50$ results (black lines). For the cumulative high-redshift optical depth,
\beq
\tau_{\rm PC}(15, 50) < 0.019 \; (95\%\; \mathrm{C.L.}),
\eeq
we also obtain consistent upper limits to the $\zmax=30$ case of Eq.~(\ref{eq:tauPC15}).

\begin{figure}[ht]
\includegraphics[width=0.48\textwidth]{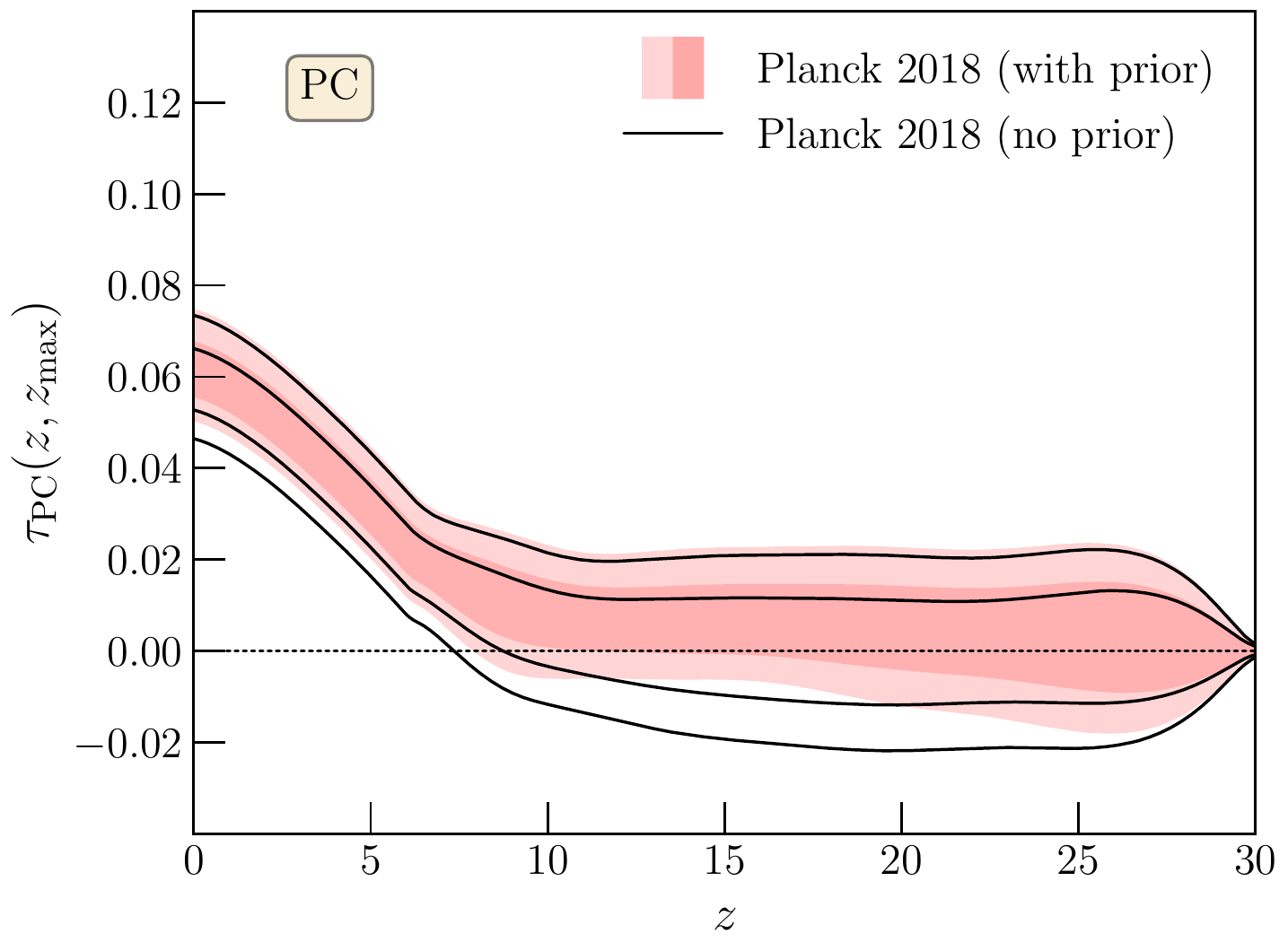}
\caption{Cumulative optical depth $\tau_{\rm PC}(z, \zmax)$ from the model-independent PC approach with and without the physicality prior.   The prior helps exclude models with negative contributions at high-$z$ but conservatively allows some negative values at $z \sim z_{\rm max}=30$ where the oscillations in the PCs no longer cancel due to the cumulative integration.  Upper limits are nearly unaffected. 
}
\label{fig:plot_taugtz_2018_with_vs_without_physicality_prior}
\end{figure}

\begin{figure}[ht]
\includegraphics[width=0.45\textwidth]{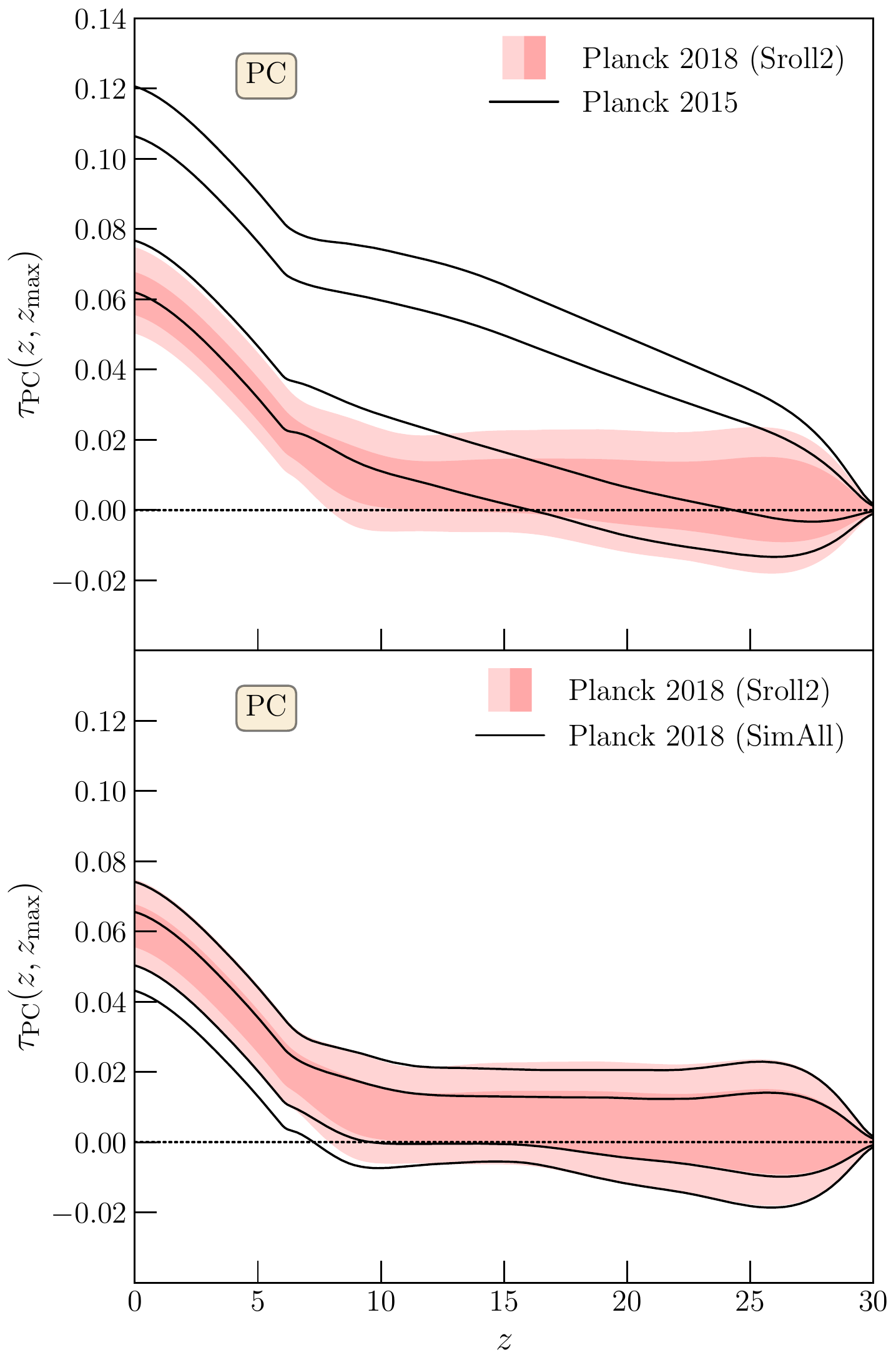}
\caption{Changes in the
cumulative optical depth $\tau_{\rm PC}(z, \zmax)$ (as in Fig.~\ref{fig:plot_taugtz_2018_with_vs_without_physicality_prior}) 
between Planck 2015 and 2018 (top panel) and between 2018 \texttt{SRoll2} (default) and $\texttt{SimAll}$.  Planck 2015 favored contributions at high $z$ which are disallowed in 2018 whereas the changes due to \texttt{SRoll2} mainly tighten the lower limits at low $z$.
}
\label{fig:plot_taugtz_2015_vs_2018_simallEE_vs_2018_srollv2}
\end{figure}

\begin{figure}[ht]
\includegraphics[width=0.48\textwidth]{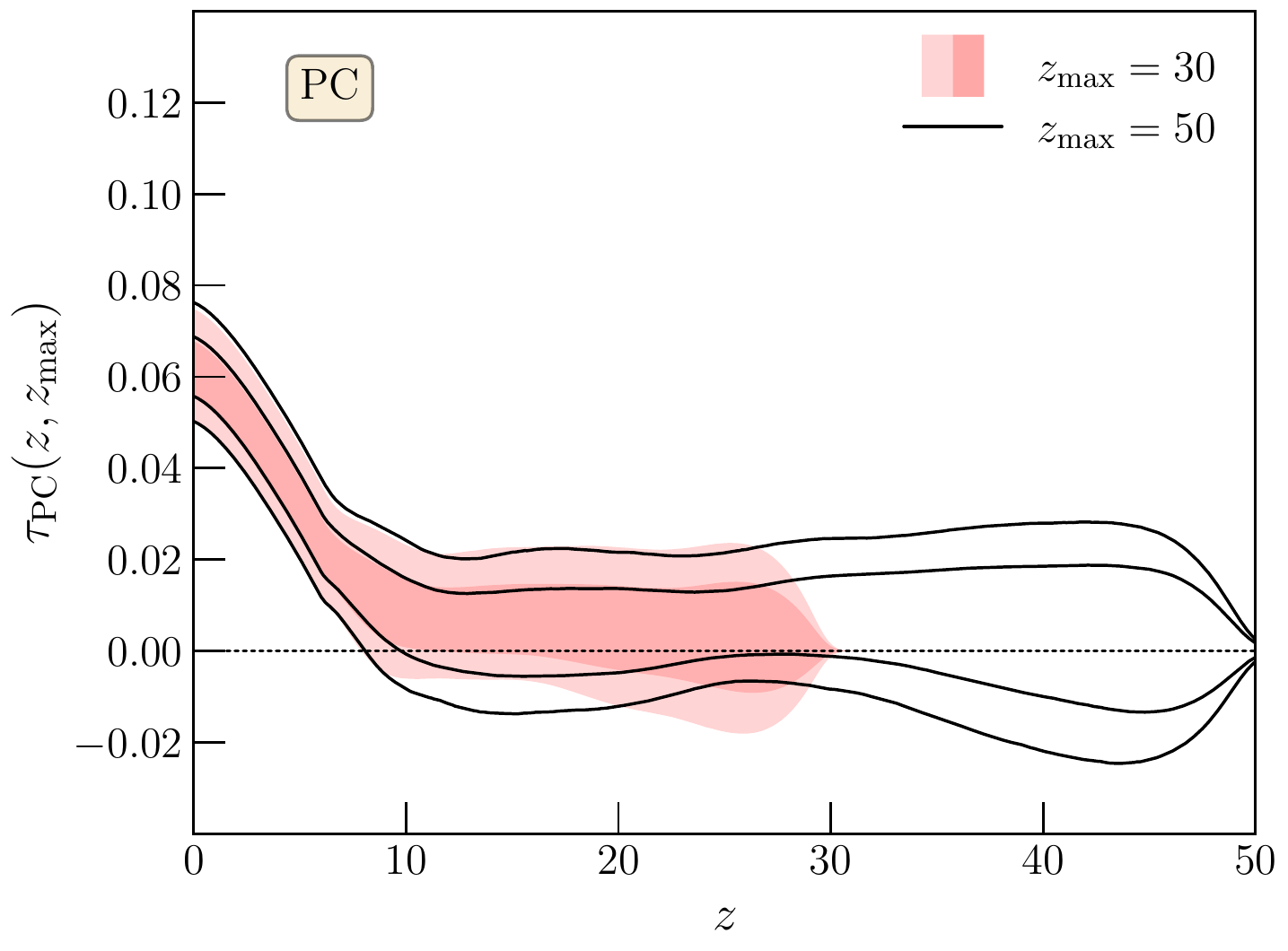}
\caption{Robustness of the cumulative optical depth $\tau_{\rm PC}(z, \zmax)$ to changes from the default $\zmax = 30$ to $\zmax = 50$. Upper limits at $z \lesssim 30$ are nearly unaffected whereas extending $\zmax$ allows the high-$z$ component to come from even higher-$z$.
}
\label{fig:plot_taugtz_zmax30_vs_zmax50}
\end{figure}

\section{Conclusion}
\label{sec:conclusion}

In conclusion, we have used constraints on reionization principal components of the large scale $EE$ polarization to produce a fast and accurate effective likelihood for the final release of Planck 2018 data.  We replaced the official $\texttt{SimAll}$ likelihood for low-$\ell$ $EE$ with the $\texttt{SRoll2}$ likelihood independently released in 2019 which uses improved foreground cleaning methods. Our effective likelihood code package is called \relike\  and is available publicly on Github.
We expect the code to facilitate the testing of \textit{any} global ionization history between $6 < z < 30$ with the Planck data and to enable fast and consistent joint analyses with other reionization datasets. 

To test this effective likelihood code, we used two examples: 1) a canonical tanh model with a single transition redshift; and 2) a two-parameter toy model consisting of two tanh transitions allowing for an additional high-redshift ionization plateau. We demonstrated accuracy of RELIKE for both examples by comparing model parameter constraints with those obtained from sampling the exact Planck likelihoods using a modified CAMB code. We also showed excellent agreement between the confidence levels of the cumulative optical depth evolution between those of the exact ionization models and their 5-PC representations.

Besides conducting explicit model testing with  \relike, we also extracted model-independent constraints using the PC chains themselves. We summarize these results below.

\begin{itemize}

    \item {We obtained the total optical depth constraint $\tau_{\rm PC} = 0.0619^{+0.0056}_{-0.0068}$ (68\% C.L.) 
    using PCs in our default likelihoods \texttt{plik\_lite\_TTTEEE+lowl+Sroll2}.}
    
    \item {Using the \texttt{SRoll2} likelihood tightened the PC constraint on $\tau$ and shifted it toward slightly higher values of $\tau$ compared to the \texttt{plik\_lite\_TTTEEE+lowl+SimAll} result $\tau_{\rm PC} = 0.0582 ^{+0.0072}_{-0.0083}$ (68\% C.L.).}
    
    \item {We checked the robustness of our $\zmax=30$ results by performing a PC analysis of the Planck data with $\zmax = 50$ where 7 PCs were used for the larger redshift range. The results changed negligibly: $\tau_{\rm PC} = 0.0626 ^{+0.0061}_{-0.0072}$ (68\% C.L.).}
    
    \item{The constraint on the high-redshift optical depth $\tau_{\rm PC}(15, 50) < 0.020$ (95\% C.L.) is also stable when extending $\zmax$ to 50: $\tau_{\rm PC}(15, 30) < 0.019$ (95\% C.L.).}
    
     \item {Our high-redshift optical depth upper limit is however a factor of $\sim3$ larger
     than that from the Planck 2018 cosmological parameter paper $\tau(15, 30) < 0.007$ (95\% C.L.) obtained from another model-independent method called FlexKnot. Using explicitly a two-step toy model (without using any PCs), we tested that $\tau_{\rm 2step}(15, 30) < 0.016$ (95\% C.L.) is still allowed, which is consistent with our PC results.}
     
\end{itemize}

Finally, we clarified the effects of applying a range-bound prior on the PC amplitude while deriving model-independent constraints on the optical depth (note that this prior does not enter the \relike\  code). We demonstrated (see Fig.~\ref{fig:plot_taugtz_2018_with_vs_without_physicality_prior}) that removing more unphysical models would only loosen the high-$z$ optical depth upper limit, in contrary to what was claimed in Ref.~\cite{Millea:2018bko}. \\

Given that the CMB intrinsically constrains integrated quantities of the ionization history, we recommend that model-independent constraints be formulated in the cumulative optical depth space rather than the ionization history space. There is also a tendancy in the literature to compare a specific model with more general constraints on $x_e(z)$. We caution the reader that these constraints are derived with a specific prior assumption and sometimes on a limited set of models. So our recommendation for constraining specific models is to use the \relike\  code which properly returns the effective Planck likelihood for any specific $x_e(z)$ function.

We also recommend using \relike\  to properly combine CMB large-scale datasets with other reionization datasets such as kSZ, galaxy luminosity functions, quasars spectra, line-intensity mapping, star formation rate, etc. In the past, joint analyses of the reionization history often used the Planck tanh constraint as a $\tau$ prior, which is an approximate proxy for the actual likelihood of the model in question. With the release of \relike, one can now easily obtain the full Planck constraint on the specific model and conduct consistent joint analyses. 

To do so, one can interface the python \relike\  package with existing MCMC samplers for generic data such as Cobaya and CosmoSIS, or those adapted for specific reionization datasets (e.g.\ 21CMMC~\cite{Greig:2015qca} and CosmoMCReion~\cite{Chatterjee:2021ygm}). When the joint dataset is constraining enough (e.g.\ kSZ), one can sample while varying cosmological parameters since the \relike\  likelihood is valid for cosmologies consistent with the Planck best-fit. We expect that joint analyses work in the future (see e.g.\ Ref.~\cite{Ahn:2020btj, Qin:2020xrg} for some recent examples) would be made much easier with \relike. \\

In the future, \relike\ can be extended in several ways.
Recent studies of quasars may indicate that the Universe have not been entirely ionized by $z = 6$ ($\sim$80\% ionized at $z = 6$)~\cite{Becker:2021jyx}. If this result is confirmed,
%in the future, 
the PC approach can be easily modified for a smaller $\zmin$ in a manner similar to the $\zmax$ study here. 

Likewise current and future CMB experiments such as CLASS~\cite{Watts:2018etg} and LiteBIRD~\cite{Hazumi:2021yqq} may also improve on Planck and eventually provide cosmic variance limited measurements of CMB polarization on the largest scales, yielding improved reionization constraints. The same approach used here can be easily updated to capture the better constraints from these data in a joint effective likelihood. \\

In sum, our release of the \relike\  code significantly reduces the time to constrain specific ionization models with Planck. It also opens the way for consistent and fast joint analyses of current and future CMB large-scale polarization data with a variety of increasingly rich reionization datasets in the next decades.

\begin{acknowledgements}
We thank Tzu-Ching Chang, GuoChao Sun, Lluis Mas-Aribas and Olivier Dor\'{e} for useful discussions. W.H. was supported by U.S. Dept. of Energy contract DE-FG02-13ER41958 and the Simons Foundation.
\end{acknowledgements}

\bibliography{rei.bib}

\begin{thebibliography}{33}%
\makeatletter
\providecommand \@ifxundefined [1]{%
 \@ifx{#1\undefined}
}%
\providecommand \@ifnum [1]{%
 \ifnum #1\expandafter \@firstoftwo
 \else \expandafter \@secondoftwo
 \fi
}%
\providecommand \@ifx [1]{%
 \ifx #1\expandafter \@firstoftwo
 \else \expandafter \@secondoftwo
 \fi
}%
\providecommand \natexlab [1]{#1}%
\providecommand \enquote  [1]{``#1''}%
\providecommand \bibnamefont  [1]{#1}%
\providecommand \bibfnamefont [1]{#1}%
\providecommand \citenamefont [1]{#1}%
\providecommand \href@noop [0]{\@secondoftwo}%
\providecommand \href [0]{\begingroup \@sanitize@url \@href}%
\providecommand \@href[1]{\@@startlink{#1}\@@href}%
\providecommand \@@href[1]{\endgroup#1\@@endlink}%
\providecommand \@sanitize@url [0]{\catcode `\\12\catcode `\$12\catcode
  `\&12\catcode `\#12\catcode `\^12\catcode `\_12\catcode `\%12\relax}%
\providecommand \@@startlink[1]{}%
\providecommand \@@endlink[0]{}%
\providecommand \url  [0]{\begingroup\@sanitize@url \@url }%
\providecommand \@url [1]{\endgroup\@href {#1}{\urlprefix }}%
\providecommand \urlprefix  [0]{URL }%
\providecommand \Eprint [0]{\href }%
\providecommand \doibase [0]{http://dx.doi.org/}%
\providecommand \selectlanguage [0]{\@gobble}%
\providecommand \bibinfo  [0]{\@secondoftwo}%
\providecommand \bibfield  [0]{\@secondoftwo}%
\providecommand \translation [1]{[#1]}%
\providecommand \BibitemOpen [0]{}%
\providecommand \bibitemStop [0]{}%
\providecommand \bibitemNoStop [0]{.\EOS\space}%
\providecommand \EOS [0]{\spacefactor3000\relax}%
\providecommand \BibitemShut  [1]{\csname bibitem#1\endcsname}%
\let\auto@bib@innerbib\@empty
%</preamble>
\bibitem [{\citenamefont {Aghanim}\ \emph {et~al.}(2018)\citenamefont {Aghanim}
  \emph {et~al.}}]{Aghanim:2018eyx}%
  \BibitemOpen
  \bibfield  {author} {\bibinfo {author} {\bibfnamefont {N.}~\bibnamefont
  {Aghanim}} \emph {et~al.} (\bibinfo {collaboration} {Planck}),\ }\href@noop
  {} {\  (\bibinfo {year} {2018})},\ \Eprint {http://arxiv.org/abs/1807.06209}
  {arXiv:1807.06209 [astro-ph.CO]} \BibitemShut {NoStop}%
\bibitem [{\citenamefont {{Mesinger}}(2016)}]{2016ASSL..423.....M}%
  \BibitemOpen
  \bibinfo {editor} {\bibfnamefont {A.}~\bibnamefont {{Mesinger}}},\ ed.,\
  \href {\doibase 10.1007/978-3-319-21957-8} {\emph {\bibinfo {title}
  {Understanding the Epoch of Cosmic Reionization: Challenges and Progress}}},\
  \bibinfo {series} {Astrophysics and Space Science Library}, Vol.\ \bibinfo
  {volume} {423}\ (\bibinfo {year} {2016})\BibitemShut {NoStop}%
\bibitem [{\citenamefont {Smith}\ \emph {et~al.}(2006)\citenamefont {Smith},
  \citenamefont {Hu},\ and\ \citenamefont {Kaplinghat}}]{Smith:2006nk}%
  \BibitemOpen
  \bibfield  {author} {\bibinfo {author} {\bibfnamefont {K.~M.}\ \bibnamefont
  {Smith}}, \bibinfo {author} {\bibfnamefont {W.}~\bibnamefont {Hu}}, \ and\
  \bibinfo {author} {\bibfnamefont {M.}~\bibnamefont {Kaplinghat}},\ }\href
  {\doibase 10.1103/PhysRevD.74.123002} {\bibfield  {journal} {\bibinfo
  {journal} {Phys. Rev.}\ }\textbf {\bibinfo {volume} {D74}},\ \bibinfo {pages}
  {123002} (\bibinfo {year} {2006})},\ \Eprint
  {http://arxiv.org/abs/astro-ph/0607315} {arXiv:astro-ph/0607315 [astro-ph]}
  \BibitemShut {NoStop}%
%%CITATION = ASTRO-PH/0607315;%%
\bibitem [{\citenamefont {Hu}\ and\ \citenamefont {Jain}(2004)}]{Hu:2003pt}%
  \BibitemOpen
  \bibfield  {author} {\bibinfo {author} {\bibfnamefont {W.}~\bibnamefont
  {Hu}}\ and\ \bibinfo {author} {\bibfnamefont {B.}~\bibnamefont {Jain}},\
  }\href {\doibase 10.1103/PhysRevD.70.043009} {\bibfield  {journal} {\bibinfo
  {journal} {Phys. Rev.}\ }\textbf {\bibinfo {volume} {D70}},\ \bibinfo {pages}
  {043009} (\bibinfo {year} {2004})},\ \Eprint
  {http://arxiv.org/abs/astro-ph/0312395} {arXiv:astro-ph/0312395 [astro-ph]}
  \BibitemShut {NoStop}%
%%CITATION = ASTRO-PH/0312395;%%
\bibitem [{\citenamefont {{McQuinn}}\ \emph {et~al.}(2005)\citenamefont
  {{McQuinn}}, \citenamefont {{Furlanetto}}, \citenamefont {{Hernquist}},
  \citenamefont {{Zahn}},\ and\ \citenamefont {{Zaldarriaga}}}]{mcquinn_2005}%
  \BibitemOpen
  \bibfield  {author} {\bibinfo {author} {\bibfnamefont {M.}~\bibnamefont
  {{McQuinn}}}, \bibinfo {author} {\bibfnamefont {S.~R.}\ \bibnamefont
  {{Furlanetto}}}, \bibinfo {author} {\bibfnamefont {L.}~\bibnamefont
  {{Hernquist}}}, \bibinfo {author} {\bibfnamefont {O.}~\bibnamefont {{Zahn}}},
  \ and\ \bibinfo {author} {\bibfnamefont {M.}~\bibnamefont {{Zaldarriaga}}},\
  }\href {\doibase 10.1086/432049} {\bibfield  {journal} {\bibinfo  {journal}
  {\apj}\ }\textbf {\bibinfo {volume} {630}},\ \bibinfo {pages} {643} (\bibinfo
  {year} {2005})},\ \Eprint {http://arxiv.org/abs/astro-ph/0504189}
  {arXiv:astro-ph/0504189 [astro-ph]} \BibitemShut {NoStop}%
\bibitem [{\citenamefont {Mesinger}\ \emph {et~al.}(2012)\citenamefont
  {Mesinger}, \citenamefont {McQuinn},\ and\ \citenamefont
  {Spergel}}]{mesinger_2012_kSZ}%
  \BibitemOpen
  \bibfield  {author} {\bibinfo {author} {\bibfnamefont {A.}~\bibnamefont
  {Mesinger}}, \bibinfo {author} {\bibfnamefont {M.}~\bibnamefont {McQuinn}}, \
  and\ \bibinfo {author} {\bibfnamefont {D.~N.}\ \bibnamefont {Spergel}},\
  }\href {\doibase 10.1111/j.1365-2966.2012.20713.x} {\bibfield  {journal}
  {\bibinfo  {journal} {Monthly Notices of the Royal Astronomical Society}\
  }\textbf {\bibinfo {volume} {422}},\ \bibinfo {pages} {1403} (\bibinfo {year}
  {2012})},\ \Eprint
  {http://arxiv.org/abs/http://oup.prod.sis.lan/mnras/article-pdf/422/2/1403/3487085/mnras0422-1403.pdf}
  {http://oup.prod.sis.lan/mnras/article-pdf/422/2/1403/3487085/mnras0422-1403.pdf}
  \BibitemShut {NoStop}%
\bibitem [{\citenamefont {Hu}\ and\ \citenamefont {Holder}(2003)}]{Hu:2003gh}%
  \BibitemOpen
  \bibfield  {author} {\bibinfo {author} {\bibfnamefont {W.}~\bibnamefont
  {Hu}}\ and\ \bibinfo {author} {\bibfnamefont {G.~P.}\ \bibnamefont
  {Holder}},\ }\href {\doibase 10.1103/PhysRevD.68.023001} {\bibfield
  {journal} {\bibinfo  {journal} {Phys. Rev.}\ }\textbf {\bibinfo {volume}
  {D68}},\ \bibinfo {pages} {023001} (\bibinfo {year} {2003})},\ \Eprint
  {http://arxiv.org/abs/astro-ph/0303400} {arXiv:astro-ph/0303400 [astro-ph]}
  \BibitemShut {NoStop}%
%%CITATION = ASTRO-PH/0303400;%%
\bibitem [{\citenamefont {Mortonson}\ and\ \citenamefont
  {Hu}(2008{\natexlab{a}})}]{Mortonson:2007hq}%
  \BibitemOpen
  \bibfield  {author} {\bibinfo {author} {\bibfnamefont {M.~J.}\ \bibnamefont
  {Mortonson}}\ and\ \bibinfo {author} {\bibfnamefont {W.}~\bibnamefont {Hu}},\
  }\href {\doibase 10.1086/523958} {\bibfield  {journal} {\bibinfo  {journal}
  {Astrophys. J.}\ }\textbf {\bibinfo {volume} {672}},\ \bibinfo {pages} {737}
  (\bibinfo {year} {2008}{\natexlab{a}})},\ \Eprint
  {http://arxiv.org/abs/0705.1132} {arXiv:0705.1132 [astro-ph]} \BibitemShut
  {NoStop}%
%%CITATION = ARXIV:0705.1132;%%
\bibitem [{\citenamefont {Mortonson}\ and\ \citenamefont
  {Hu}(2008{\natexlab{b}})}]{Mortonson:2008rx}%
  \BibitemOpen
  \bibfield  {author} {\bibinfo {author} {\bibfnamefont {M.~J.}\ \bibnamefont
  {Mortonson}}\ and\ \bibinfo {author} {\bibfnamefont {W.}~\bibnamefont {Hu}},\
  }\href {\doibase 10.1086/593031} {\bibfield  {journal} {\bibinfo  {journal}
  {Astrophys. J.}\ }\textbf {\bibinfo {volume} {686}},\ \bibinfo {pages} {L53}
  (\bibinfo {year} {2008}{\natexlab{b}})},\ \Eprint
  {http://arxiv.org/abs/0804.2631} {arXiv:0804.2631 [astro-ph]} \BibitemShut
  {NoStop}%
%%CITATION = ARXIV:0804.2631;%%
\bibitem [{\citenamefont {Heinrich}\ \emph {et~al.}(2017)\citenamefont
  {Heinrich}, \citenamefont {Miranda},\ and\ \citenamefont
  {Hu}}]{Heinrich:2016ojb}%
  \BibitemOpen
  \bibfield  {author} {\bibinfo {author} {\bibfnamefont {C.~H.}\ \bibnamefont
  {Heinrich}}, \bibinfo {author} {\bibfnamefont {V.}~\bibnamefont {Miranda}}, \
  and\ \bibinfo {author} {\bibfnamefont {W.}~\bibnamefont {Hu}},\ }\href
  {\doibase 10.1103/PhysRevD.95.023513} {\bibfield  {journal} {\bibinfo
  {journal} {Phys. Rev.}\ }\textbf {\bibinfo {volume} {D95}},\ \bibinfo {pages}
  {023513} (\bibinfo {year} {2017})},\ \Eprint
  {http://arxiv.org/abs/1609.04788} {arXiv:1609.04788 [astro-ph.CO]}
  \BibitemShut {NoStop}%
%%CITATION = ARXIV:1609.04788;%%
\bibitem [{\citenamefont {Ade}\ \emph {et~al.}(2014)\citenamefont {Ade} \emph
  {et~al.}}]{Planck:2013jfk}%
  \BibitemOpen
  \bibfield  {author} {\bibinfo {author} {\bibfnamefont {P.~A.~R.}\
  \bibnamefont {Ade}} \emph {et~al.} (\bibinfo {collaboration} {Planck}),\
  }\href {\doibase 10.1051/0004-6361/201321569} {\bibfield  {journal} {\bibinfo
   {journal} {Astron. Astrophys.}\ }\textbf {\bibinfo {volume} {571}},\
  \bibinfo {pages} {A22} (\bibinfo {year} {2014})},\ \Eprint
  {http://arxiv.org/abs/1303.5082} {arXiv:1303.5082 [astro-ph.CO]} \BibitemShut
  {NoStop}%
%%CITATION = ARXIV:1303.5082;%%
\bibitem [{\citenamefont {Obied}\ \emph {et~al.}(2018)\citenamefont {Obied},
  \citenamefont {Dvorkin}, \citenamefont {Heinrich}, \citenamefont {Hu},\ and\
  \citenamefont {Miranda}}]{Obied:2018qdr}%
  \BibitemOpen
  \bibfield  {author} {\bibinfo {author} {\bibfnamefont {G.}~\bibnamefont
  {Obied}}, \bibinfo {author} {\bibfnamefont {C.}~\bibnamefont {Dvorkin}},
  \bibinfo {author} {\bibfnamefont {C.}~\bibnamefont {Heinrich}}, \bibinfo
  {author} {\bibfnamefont {W.}~\bibnamefont {Hu}}, \ and\ \bibinfo {author}
  {\bibfnamefont {V.}~\bibnamefont {Miranda}},\ }\href@noop {} {\  (\bibinfo
  {year} {2018})},\ \Eprint {http://arxiv.org/abs/1803.01858} {arXiv:1803.01858
  [astro-ph.CO]} \BibitemShut {NoStop}%
%%CITATION = ARXIV:1803.01858;%%
\bibitem [{\citenamefont {Dai}\ \emph {et~al.}(2015)\citenamefont {Dai},
  \citenamefont {Guo},\ and\ \citenamefont {Cai}}]{Dai:2015dwa}%
  \BibitemOpen
  \bibfield  {author} {\bibinfo {author} {\bibfnamefont {W.-M.}\ \bibnamefont
  {Dai}}, \bibinfo {author} {\bibfnamefont {Z.-K.}\ \bibnamefont {Guo}}, \ and\
  \bibinfo {author} {\bibfnamefont {R.-G.}\ \bibnamefont {Cai}},\ }\href
  {\doibase 10.1103/PhysRevD.92.123521} {\bibfield  {journal} {\bibinfo
  {journal} {Phys. Rev.}\ }\textbf {\bibinfo {volume} {D92}},\ \bibinfo {pages}
  {123521} (\bibinfo {year} {2015})},\ \Eprint
  {http://arxiv.org/abs/1509.01501} {arXiv:1509.01501 [astro-ph.CO]}
  \BibitemShut {NoStop}%
%%CITATION = ARXIV:1509.01501;%%
\bibitem [{\citenamefont {Millea}\ and\ \citenamefont
  {Bouchet}(2018)}]{Millea:2018bko}%
  \BibitemOpen
  \bibfield  {author} {\bibinfo {author} {\bibfnamefont {M.}~\bibnamefont
  {Millea}}\ and\ \bibinfo {author} {\bibfnamefont {F.}~\bibnamefont
  {Bouchet}},\ }\href@noop {} {\  (\bibinfo {year} {2018})},\ \Eprint
  {http://arxiv.org/abs/1804.08476} {arXiv:1804.08476 [astro-ph.CO]}
  \BibitemShut {NoStop}%
%%CITATION = ARXIV:1804.08476;%%
\bibitem [{\citenamefont {Heinrich}\ and\ \citenamefont
  {Hu}(2018)}]{Heinrich:2018btc}%
  \BibitemOpen
  \bibfield  {author} {\bibinfo {author} {\bibfnamefont {C.}~\bibnamefont
  {Heinrich}}\ and\ \bibinfo {author} {\bibfnamefont {W.}~\bibnamefont {Hu}},\
  }\href {\doibase 10.1103/PhysRevD.98.063514} {\bibfield  {journal} {\bibinfo
  {journal} {Phys. Rev. D}\ }\textbf {\bibinfo {volume} {98}},\ \bibinfo
  {pages} {063514} (\bibinfo {year} {2018})},\ \Eprint
  {http://arxiv.org/abs/1802.00791} {arXiv:1802.00791 [astro-ph.CO]}
  \BibitemShut {NoStop}%
\bibitem [{\citenamefont {Delouis}\ \emph {et~al.}(2019)\citenamefont
  {Delouis}, \citenamefont {Pagano}, \citenamefont {Mottet}, \citenamefont
  {Puget},\ and\ \citenamefont {Vibert}}]{Delouis:2019bub}%
  \BibitemOpen
  \bibfield  {author} {\bibinfo {author} {\bibfnamefont {J.~M.}\ \bibnamefont
  {Delouis}}, \bibinfo {author} {\bibfnamefont {L.}~\bibnamefont {Pagano}},
  \bibinfo {author} {\bibfnamefont {S.}~\bibnamefont {Mottet}}, \bibinfo
  {author} {\bibfnamefont {J.~L.}\ \bibnamefont {Puget}}, \ and\ \bibinfo
  {author} {\bibfnamefont {L.}~\bibnamefont {Vibert}},\ }\href {\doibase
  10.1051/0004-6361/201834882} {\bibfield  {journal} {\bibinfo  {journal}
  {Astron. Astrophys.}\ }\textbf {\bibinfo {volume} {629}},\ \bibinfo {pages}
  {A38} (\bibinfo {year} {2019})},\ \Eprint {http://arxiv.org/abs/1901.11386}
  {arXiv:1901.11386 [astro-ph.CO]} \BibitemShut {NoStop}%
\bibitem [{\citenamefont {Becker}\ \emph {et~al.}(2015)\citenamefont {Becker},
  \citenamefont {Bolton},\ and\ \citenamefont {Lidz}}]{Becker:2015lua}%
  \BibitemOpen
  \bibfield  {author} {\bibinfo {author} {\bibfnamefont {G.~D.}\ \bibnamefont
  {Becker}}, \bibinfo {author} {\bibfnamefont {J.~S.}\ \bibnamefont {Bolton}},
  \ and\ \bibinfo {author} {\bibfnamefont {A.}~\bibnamefont {Lidz}},\ }\href
  {\doibase 10.1017/pasa.2015.45} {\bibfield  {journal} {\bibinfo  {journal}
  {Publ. Astron. Soc. Austral.}\ }\textbf {\bibinfo {volume} {32}},\ \bibinfo
  {pages} {45} (\bibinfo {year} {2015})},\ \Eprint
  {http://arxiv.org/abs/1510.03368} {arXiv:1510.03368 [astro-ph.CO]}
  \BibitemShut {NoStop}%
%%CITATION = ARXIV:1510.03368;%%
\bibitem [{\citenamefont {Lewis}\ \emph {et~al.}(2000)\citenamefont {Lewis},
  \citenamefont {Challinor},\ and\ \citenamefont {Lasenby}}]{Lewis:1999bs}%
  \BibitemOpen
  \bibfield  {author} {\bibinfo {author} {\bibfnamefont {A.}~\bibnamefont
  {Lewis}}, \bibinfo {author} {\bibfnamefont {A.}~\bibnamefont {Challinor}}, \
  and\ \bibinfo {author} {\bibfnamefont {A.}~\bibnamefont {Lasenby}},\ }\href
  {\doibase 10.1086/309179} {\bibfield  {journal} {\bibinfo  {journal}
  {Astrophys. J.}\ }\textbf {\bibinfo {volume} {538}},\ \bibinfo {pages} {473}
  (\bibinfo {year} {2000})},\ \Eprint {http://arxiv.org/abs/astro-ph/9911177}
  {arXiv:astro-ph/9911177 [astro-ph]} \BibitemShut {NoStop}%
%%CITATION = ASTRO-PH/9911177;%%
\bibitem [{\citenamefont {Howlett}\ \emph {et~al.}(2012)\citenamefont
  {Howlett}, \citenamefont {Lewis}, \citenamefont {Hall},\ and\ \citenamefont
  {Challinor}}]{Howlett:2012mh}%
  \BibitemOpen
  \bibfield  {author} {\bibinfo {author} {\bibfnamefont {C.}~\bibnamefont
  {Howlett}}, \bibinfo {author} {\bibfnamefont {A.}~\bibnamefont {Lewis}},
  \bibinfo {author} {\bibfnamefont {A.}~\bibnamefont {Hall}}, \ and\ \bibinfo
  {author} {\bibfnamefont {A.}~\bibnamefont {Challinor}},\ }\href {\doibase
  10.1088/1475-7516/2012/04/027} {\bibfield  {journal} {\bibinfo  {journal}
  {JCAP}\ }\textbf {\bibinfo {volume} {1204}},\ \bibinfo {pages} {027}
  (\bibinfo {year} {2012})},\ \Eprint {http://arxiv.org/abs/1201.3654}
  {arXiv:1201.3654 [astro-ph.CO]} \BibitemShut {NoStop}%
%%CITATION = ARXIV:1201.3654;%%
\bibitem [{\citenamefont {Becker}\ \emph {et~al.}(2011)\citenamefont {Becker},
  \citenamefont {Bolton}, \citenamefont {Haehnelt},\ and\ \citenamefont
  {Sargent}}]{Becker:2010cu}%
  \BibitemOpen
  \bibfield  {author} {\bibinfo {author} {\bibfnamefont {G.~D.}\ \bibnamefont
  {Becker}}, \bibinfo {author} {\bibfnamefont {J.~S.}\ \bibnamefont {Bolton}},
  \bibinfo {author} {\bibfnamefont {M.~G.}\ \bibnamefont {Haehnelt}}, \ and\
  \bibinfo {author} {\bibfnamefont {W.~L.~W.}\ \bibnamefont {Sargent}},\ }\href
  {\doibase 10.1111/j.1365-2966.2010.17507.x} {\bibfield  {journal} {\bibinfo
  {journal} {Mon. Not. Roy. Astron. Soc.}\ }\textbf {\bibinfo {volume} {410}},\
  \bibinfo {pages} {1096} (\bibinfo {year} {2011})},\ \Eprint
  {http://arxiv.org/abs/1008.2622} {arXiv:1008.2622 [astro-ph.CO]} \BibitemShut
  {NoStop}%
%%CITATION = ARXIV:1008.2622;%%
\bibitem [{\citenamefont {Aghanim}\ \emph {et~al.}(2020)\citenamefont {Aghanim}
  \emph {et~al.}}]{Aghanim:2019ame}%
  \BibitemOpen
  \bibfield  {author} {\bibinfo {author} {\bibfnamefont {N.}~\bibnamefont
  {Aghanim}} \emph {et~al.} (\bibinfo {collaboration} {Planck}),\ }\href
  {\doibase 10.1051/0004-6361/201936386} {\bibfield  {journal} {\bibinfo
  {journal} {Astron. Astrophys.}\ }\textbf {\bibinfo {volume} {641}},\ \bibinfo
  {pages} {A5} (\bibinfo {year} {2020})},\ \Eprint
  {http://arxiv.org/abs/1907.12875} {arXiv:1907.12875 [astro-ph.CO]}
  \BibitemShut {NoStop}%
\bibitem [{\citenamefont {Torrado}\ and\ \citenamefont
  {Lewis}(2020)}]{Torrado:2020dgo}%
  \BibitemOpen
  \bibfield  {author} {\bibinfo {author} {\bibfnamefont {J.}~\bibnamefont
  {Torrado}}\ and\ \bibinfo {author} {\bibfnamefont {A.}~\bibnamefont
  {Lewis}},\ }\href@noop {} {\  (\bibinfo {year} {2020})},\ \Eprint
  {http://arxiv.org/abs/2005.05290} {arXiv:2005.05290 [astro-ph.IM]}
  \BibitemShut {NoStop}%
\bibitem [{\citenamefont {{Torrado}}\ and\ \citenamefont
  {{Lewis}}(2020)}]{2020arXiv200505290T}%
  \BibitemOpen
  \bibfield  {author} {\bibinfo {author} {\bibfnamefont {J.}~\bibnamefont
  {{Torrado}}}\ and\ \bibinfo {author} {\bibfnamefont {A.}~\bibnamefont
  {{Lewis}}},\ }\href@noop {} {\bibfield  {journal} {\bibinfo  {journal} {arXiv
  e-prints}\ ,\ \bibinfo {eid} {arXiv:2005.05290}} (\bibinfo {year} {2020})},\
  \Eprint {http://arxiv.org/abs/2005.05290} {arXiv:2005.05290 [astro-ph.IM]}
  \BibitemShut {NoStop}%
\bibitem [{\citenamefont {Zuntz}\ \emph {et~al.}(2015)\citenamefont {Zuntz},
  \citenamefont {Paterno}, \citenamefont {Jennings}, \citenamefont {Rudd},
  \citenamefont {Manzotti}, \citenamefont {Dodelson}, \citenamefont {Bridle},
  \citenamefont {Sehrish},\ and\ \citenamefont {Kowalkowski}}]{Zuntz:2014csq}%
  \BibitemOpen
  \bibfield  {author} {\bibinfo {author} {\bibfnamefont {J.}~\bibnamefont
  {Zuntz}}, \bibinfo {author} {\bibfnamefont {M.}~\bibnamefont {Paterno}},
  \bibinfo {author} {\bibfnamefont {E.}~\bibnamefont {Jennings}}, \bibinfo
  {author} {\bibfnamefont {D.}~\bibnamefont {Rudd}}, \bibinfo {author}
  {\bibfnamefont {A.}~\bibnamefont {Manzotti}}, \bibinfo {author}
  {\bibfnamefont {S.}~\bibnamefont {Dodelson}}, \bibinfo {author}
  {\bibfnamefont {S.}~\bibnamefont {Bridle}}, \bibinfo {author} {\bibfnamefont
  {S.}~\bibnamefont {Sehrish}}, \ and\ \bibinfo {author} {\bibfnamefont
  {J.}~\bibnamefont {Kowalkowski}},\ }\href {\doibase
  10.1016/j.ascom.2015.05.005} {\bibfield  {journal} {\bibinfo  {journal}
  {Astron. Comput.}\ }\textbf {\bibinfo {volume} {12}},\ \bibinfo {pages} {45}
  (\bibinfo {year} {2015})},\ \Eprint {http://arxiv.org/abs/1409.3409}
  {arXiv:1409.3409 [astro-ph.CO]} \BibitemShut {NoStop}%
\bibitem [{\citenamefont {Pagano}\ \emph {et~al.}(2020)\citenamefont {Pagano},
  \citenamefont {Delouis}, \citenamefont {Mottet}, \citenamefont {Puget},\ and\
  \citenamefont {Vibert}}]{Pagano:2019tci}%
  \BibitemOpen
  \bibfield  {author} {\bibinfo {author} {\bibfnamefont {L.}~\bibnamefont
  {Pagano}}, \bibinfo {author} {\bibfnamefont {J.~M.}\ \bibnamefont {Delouis}},
  \bibinfo {author} {\bibfnamefont {S.}~\bibnamefont {Mottet}}, \bibinfo
  {author} {\bibfnamefont {J.~L.}\ \bibnamefont {Puget}}, \ and\ \bibinfo
  {author} {\bibfnamefont {L.}~\bibnamefont {Vibert}},\ }\href {\doibase
  10.1051/0004-6361/201936630} {\bibfield  {journal} {\bibinfo  {journal}
  {Astron. Astrophys.}\ }\textbf {\bibinfo {volume} {635}},\ \bibinfo {pages}
  {A99} (\bibinfo {year} {2020})},\ \Eprint {http://arxiv.org/abs/1908.09856}
  {arXiv:1908.09856 [astro-ph.CO]} \BibitemShut {NoStop}%
\bibitem [{\citenamefont {Ahn}\ and\ \citenamefont
  {Shapiro}(2020)}]{Ahn:2020btj}%
  \BibitemOpen
  \bibfield  {author} {\bibinfo {author} {\bibfnamefont {K.}~\bibnamefont
  {Ahn}}\ and\ \bibinfo {author} {\bibfnamefont {P.~R.}\ \bibnamefont
  {Shapiro}},\ }\href@noop {} {\  (\bibinfo {year} {2020})},\ \Eprint
  {http://arxiv.org/abs/2011.03582} {arXiv:2011.03582 [astro-ph.CO]}
  \BibitemShut {NoStop}%
\bibitem [{\citenamefont {Paoletti}\ \emph {et~al.}(2020)\citenamefont
  {Paoletti}, \citenamefont {Hazra}, \citenamefont {Finelli},\ and\
  \citenamefont {Smoot}}]{Paoletti:2020ndu}%
  \BibitemOpen
  \bibfield  {author} {\bibinfo {author} {\bibfnamefont {D.}~\bibnamefont
  {Paoletti}}, \bibinfo {author} {\bibfnamefont {D.~K.}\ \bibnamefont {Hazra}},
  \bibinfo {author} {\bibfnamefont {F.}~\bibnamefont {Finelli}}, \ and\
  \bibinfo {author} {\bibfnamefont {G.~F.}\ \bibnamefont {Smoot}},\ }\href@noop
  {} {\  (\bibinfo {year} {2020})},\ \Eprint {http://arxiv.org/abs/2005.12222}
  {arXiv:2005.12222 [astro-ph.CO]} \BibitemShut {NoStop}%
\bibitem [{\citenamefont {Greig}\ and\ \citenamefont
  {Mesinger}(2015)}]{Greig:2015qca}%
  \BibitemOpen
  \bibfield  {author} {\bibinfo {author} {\bibfnamefont {B.}~\bibnamefont
  {Greig}}\ and\ \bibinfo {author} {\bibfnamefont {A.}~\bibnamefont
  {Mesinger}},\ }\href {\doibase 10.1093/mnras/stv571} {\bibfield  {journal}
  {\bibinfo  {journal} {Mon. Not. Roy. Astron. Soc.}\ }\textbf {\bibinfo
  {volume} {449}},\ \bibinfo {pages} {4246} (\bibinfo {year} {2015})},\ \Eprint
  {http://arxiv.org/abs/1501.06576} {arXiv:1501.06576 [astro-ph.CO]}
  \BibitemShut {NoStop}%
\bibitem [{\citenamefont {Chatterjee}\ \emph {et~al.}(2021)\citenamefont
  {Chatterjee}, \citenamefont {Choudhury},\ and\ \citenamefont
  {Mitra}}]{Chatterjee:2021ygm}%
  \BibitemOpen
  \bibfield  {author} {\bibinfo {author} {\bibfnamefont {A.}~\bibnamefont
  {Chatterjee}}, \bibinfo {author} {\bibfnamefont {T.~R.}\ \bibnamefont
  {Choudhury}}, \ and\ \bibinfo {author} {\bibfnamefont {S.}~\bibnamefont
  {Mitra}},\ }\href@noop {} {\  (\bibinfo {year} {2021})},\ \Eprint
  {http://arxiv.org/abs/2101.11088} {arXiv:2101.11088 [astro-ph.CO]}
  \BibitemShut {NoStop}%
\bibitem [{\citenamefont {Qin}\ \emph {et~al.}(2020)\citenamefont {Qin},
  \citenamefont {Poulin}, \citenamefont {Mesinger}, \citenamefont {Greig},
  \citenamefont {Murray},\ and\ \citenamefont {Park}}]{Qin:2020xrg}%
  \BibitemOpen
  \bibfield  {author} {\bibinfo {author} {\bibfnamefont {Y.}~\bibnamefont
  {Qin}}, \bibinfo {author} {\bibfnamefont {V.}~\bibnamefont {Poulin}},
  \bibinfo {author} {\bibfnamefont {A.}~\bibnamefont {Mesinger}}, \bibinfo
  {author} {\bibfnamefont {B.}~\bibnamefont {Greig}}, \bibinfo {author}
  {\bibfnamefont {S.}~\bibnamefont {Murray}}, \ and\ \bibinfo {author}
  {\bibfnamefont {J.}~\bibnamefont {Park}},\ }\href@noop {} {\  (\bibinfo
  {year} {2020})},\ \Eprint {http://arxiv.org/abs/2006.16828} {arXiv:2006.16828
  [astro-ph.CO]} \BibitemShut {NoStop}%
\bibitem [{\citenamefont {Becker}\ \emph {et~al.}(2021)\citenamefont {Becker},
  \citenamefont {D'aloisio}, \citenamefont {Christenson}, \citenamefont {Zhu},
  \citenamefont {Worseck},\ and\ \citenamefont {Bolton}}]{Becker:2021jyx}%
  \BibitemOpen
  \bibfield  {author} {\bibinfo {author} {\bibfnamefont {G.~D.}\ \bibnamefont
  {Becker}}, \bibinfo {author} {\bibfnamefont {A.}~\bibnamefont {D'aloisio}},
  \bibinfo {author} {\bibfnamefont {H.~M.}\ \bibnamefont {Christenson}},
  \bibinfo {author} {\bibfnamefont {Y.}~\bibnamefont {Zhu}}, \bibinfo {author}
  {\bibfnamefont {G.}~\bibnamefont {Worseck}}, \ and\ \bibinfo {author}
  {\bibfnamefont {J.~S.}\ \bibnamefont {Bolton}},\ }\href@noop {} {\  (\bibinfo
  {year} {2021})},\ \Eprint {http://arxiv.org/abs/2103.16610} {arXiv:2103.16610
  [astro-ph.CO]} \BibitemShut {NoStop}%
\bibitem [{\citenamefont {Watts}\ \emph {et~al.}(2018)\citenamefont {Watts}
  \emph {et~al.}}]{Watts:2018etg}%
  \BibitemOpen
  \bibfield  {author} {\bibinfo {author} {\bibfnamefont {D.~J.}\ \bibnamefont
  {Watts}} \emph {et~al.},\ }\href@noop {} {\  (\bibinfo {year} {2018})},\
  \Eprint {http://arxiv.org/abs/1801.01481} {arXiv:1801.01481 [astro-ph.CO]}
  \BibitemShut {NoStop}%
%%CITATION = ARXIV:1801.01481;%%
\bibitem [{\citenamefont {Hazumi}\ \emph {et~al.}(2020)\citenamefont {Hazumi}
  \emph {et~al.}}]{Hazumi:2021yqq}%
  \BibitemOpen
  \bibfield  {author} {\bibinfo {author} {\bibfnamefont {M.}~\bibnamefont
  {Hazumi}} \emph {et~al.} (\bibinfo {collaboration} {LiteBIRD}),\ }\href
  {\doibase 10.1117/12.2563050} {\bibfield  {journal} {\bibinfo  {journal}
  {Proc. SPIE Int. Soc. Opt. Eng.}\ }\textbf {\bibinfo {volume} {11443}},\
  \bibinfo {pages} {114432F} (\bibinfo {year} {2020})},\ \Eprint
  {http://arxiv.org/abs/2101.12449} {arXiv:2101.12449 [astro-ph.IM]}
  \BibitemShut {NoStop}%
\end{thebibliography}%

\appendix

\end{document}